\documentclass[11pt, letterpaper]{article}

\pdfoutput=1
\usepackage{arxiv}
\usepackage{verbatim}

\usepackage[utf8]{inputenc} 
\usepackage[T1]{fontenc}    
\usepackage{hyperref}       
\usepackage{url}            
\usepackage{booktabs}       
\usepackage{amsfonts}       
\usepackage{nicefrac}       
\usepackage{microtype}      
\usepackage{lipsum}

\usepackage{fullpage}
\usepackage{graphicx}
\usepackage{subcaption}
\usepackage{extarrows}
\usepackage{mathtools}
\usepackage{amssymb,amsmath}
\usepackage{mathabx}
\usepackage{plain}
\usepackage{algorithm,algcompatible}
\usepackage{amsthm}
\usepackage{xcolor}
\usepackage{booktabs}

\theoremstyle{definition}
\newtheorem{definition}{Definition}[section]

\algnewcommand\INPUT{\item[\textbf{Input:}]}%
\algnewcommand\OUTPUT{\item[\textbf{Output:}]}%

\def\real{\mathop{\rm I\!R}\nolimits}

\graphicspath{{./fig/}}

\title{Finding Acceptable Parameter Regions of Stochastic Hill functions for Multisite Phosphorylation Mechanism}

\author{
  Minghan Chen\\
  Department of Computer Science\\
  Virginia Tech\\
  Blacksburg, VA 24061 \\
  \texttt{chenm@wfu.edu} \\
   \And
    Mansooreh Ahmadian\\
    Department of Computer Science\\
  Virginia Tech\\
  Blacksburg, VA 24061 \\
   \texttt{amadian@vt.edu}\\
    \And
    Layne Watson\\
    Department of Computer Science\\
  Virginia Tech\\
  Blacksburg, VA 24061 \\
    \texttt{ltw@cs.vt.edu}\\
    \And
    Yang Cao\\
    Department of Computer Science\\
  Virginia Tech\\
  Blacksburg, VA 24061 \\
    \texttt{ycao@vt.edu}\\
}

\date{}

\begin{document}

\maketitle
\thispagestyle{empty}

\begin{abstract}
Multisite phosphorylation plays an important role in regulating switchlike
protein activity and has been used widely in mathematical models. With
the development of new experimental techniques and more molecular data,
molecular phosphorylation processes emerge in many systems with increasing
complexity and sizes. These developments call for simple yet valid
stochastic models to describe various multisite phosphorylation processes,
especially in large and complex biochemical networks. To reduce model
complexity, this work aims to simplify the multisite phosphorylation
mechanism by a stochastic Hill function model. Further, this work optimizes
regions of parameter space to match simulation results from the stochastic
Hill function with the distributive multisite phosphorylation process.
While traditional parameter optimization methods have been focusing on
finding the best parameter vector, in most circumstances modelers would
like to find a set of parameter vectors that generate similar system
dynamics and results. This paper proposes a general
$\alpha$-$\beta$-$\gamma$ rule to return an acceptable parameter region
of the stochastic Hill function based on a quasi-Newton stochastic
optimization (QNSTOP) algorithm. Different objective functions are
investigated characterizing different features of the simulation-based
empirical data, among which the approximate maximum log-likelihood method
is recommended for general applications. Numerical results demonstrate
that with an appropriate parameter vector value, the stochastic Hill
function model depicts the multisite phosphorylation process well except
the initial (transient) period.
 \end{abstract}

\section{Introduction}\label{sec:introduction}
Protein phosphorylation is a fundamental molecular mechanism that regulates
protein function through the addition of a phosphate group. Many proteins
have multiple phosphorylation sites and the phosphorylation process often
targets multiple distinct sites, referred to as multisite phosphorylation.
Multisite phosphorylation facilitates the complexity and scope of
protein activity and is ubiquitous in biological processes such as cell
cycle regulation~\cite{ali2011cell}. Over the years, researchers
have revealed that multisite phosphorylation forms the basis of switchlike
response~\cite{ferrell2014ultrasensitivity} and can exhibit bistability
and periodic
oscillations~\cite{kapuy2009bistability,rubinstein2016long,barik2010model}.

Based on the enzyme (kinase) processivity, there are two categories of
multisite phosphorylation mechanisms~\cite{salazar2009multisite}. The
processive mechanism occurs when the enzyme phosphorylates all sites without
dissociation from the substrate, while in the distributive mechanism, the
enzyme dissociates from its substrate after one site becomes phosphorylated.
This paper considers the multisite phosphorylation process with an ordered
distributive mechanism, shown by
 \begin{equation}\label{sys:multi}
{\rm B}_0 \xrightleftharpoons[b_1]{f_1[ {\rm A}]}  {\rm B}_1
\xrightleftharpoons[b_2]{f_2[{\rm A}]} {\rm  B}_2 \ldots
\xrightleftharpoons[b_{n-1}]{f_{n-1}[{\rm A}]}  {\rm B}_{n-1}
\xrightleftharpoons[b_{n}]{f_{n}[{\rm A}]}  {\rm B}_n.
 \end{equation}
B is a protein (substrate) with $n$ phosphorylation sites. A is the enzyme
(kinase) that can bind to substrate B. ${\rm B}_0$ represents the
unphosphorylated form (unbound to any phosphoryl group), and ${\rm B}_n$
is the fully phosphorylated form. ${\rm B}_n$ is formed in sequential
steps, with intermediate stages increasing by one site phosphorylated until
all $n$ sites are phosphorylated. Assume that there are enough phosphate
groups for the phosphorylation process in system~\eqref{sys:multi} and all
binding sites are the same (no preference over the selection of binding
sites). Enzyme A controls the phosphorylation of ${\rm B}_i$ with the
phosphorylation activity $f_i[\rm A]$, where $f_i$ is the phosphorylation
rate and $[\rm A]$ denotes the quantity (population or concentration) of
the enzyme A. In the stochastic regime, [A] represents the enzyme population.
$b_i$ $(i=1$, $2$, $\ldots$, $n)$ denotes the dephosphorylation activity
of ${\rm B}_i$, where $b_i={k_i}[{\rm phosphotase}]$. Assuming that
the phosphotase level is constant in the enzyme-substrate
system~\eqref{sys:multi}, one can simply use $b_i$ to represent the
product of dephosphorylation rate $k_i$ and phosphotase level.

The kinetic behavior of multisite phosphorylation can be mathematically
modeled by a set of differential equations based on a series of elementary
reactions describing the enzyme-substrate binding and catalysis process.
Thus, the concentrations of all possible phosphorylated forms ${\rm[B}_i]$
is controlled by
 \begin{equation}\label{eq:multi}
\left\{ \begin{aligned}
   & \frac{ d[{\rm B}_0]}{d t} =  b_1[{\rm B}_1] -
   f_1[{\rm A}][{\rm B}_0], \\
   & \frac{ d [{\rm B}_i]}{d t} =  f_i[{\rm A}][{\rm B}_{i-1}] +
   b_{i+1}[{\rm B}_{i+1}] - (f_{i+1}[{\rm A}]+b_i)[{\rm B}_i], \text{ for }
   0 < i < n, \\
   & \frac{ d[{\rm B}_n]}{d t} =  f_n[{\rm A}][{\rm B}_{n-1}]
   - b_n[{\rm B}_n]. \\
\end{aligned} \right.
 \end{equation}
The above differential equations are a traditional deterministic method
for modeling and simulating the multisite phosphorylation. While in the
stochastic regime, the enzyme-substrate system is governed by a list of
reaction propensities, which describe the probabilities of reaction events.
Below are the reactions and corresponding propensities for the multisite
phosphorylation system~\eqref{sys:multi}, where $0 < i \le n$.

\begin{table}[ht!]
\centering
\caption{Chemical reaction and propensity calculation}
\begin{tabular}{ c c } 
\toprule
 Reaction & Propensity \\ 
 \midrule
${\rm B}_{i-1} \xrightarrow{f_{i}[{\rm A}]}  {\rm B}_i$
& $a = f_{i} [{\rm A}][{\rm B}_{i-1}]$\\ 
${\rm B}_i \xrightarrow{~~b_i~~} {\rm B}_{i-1}$ & $a = b_{i}[{\rm B}_i]$ \\
\bottomrule
\end{tabular}
\end{table}

In the above chemical reactions, if we treat [A] as a constant (parameter), the system~\eqref{sys:multi} becomes a 
linear reaction network. There are many studies on the stochastic kinetics and efficient modeling and simulation 
of such chemical reaction networks~\cite{brown2003single, Chen2019, darvey1966stochastic,gadgil2005stochastic, lente2012stochastic}. 
For example, Gadgila et al.~\cite{gadgil2005stochastic}
formulated the governing master equations of first order reactions and 
results showed that the distribution of all the system components in an 
open system follows Poisson distribution at steady state.
Lente~\cite{lente2012stochastic} developed a stochastic mapping for first order reaction networks to 
help evaluate the appropriate method between stochastic and deterministic approaches.
Our recent work~\cite{Chen2019} also presented a numerical analysis for the accuracy of the hybrid ODE/SSA method on linearly reacting systems.

However, for complex biochemical networks involving multisite protein
phosphorylation, the above kinetic system in both deterministic and
stochastic regimes increases the model size and complexity especially when
$n$ is large (a protein with $n$ phosphorylation sites has potentially $n$
to $2^n$ distinct phosphorylation forms.) On the other hand, when trying
to formulate a realistic multisite phosphorylation system, little is known
beyond specific enzyme-substrate systems. Usually, most reaction rates
are tuned to match with empirical observations and assumptions are made
with limited biological justification. Considering the complexity and lack
of knowledge, this paper proposes to model the multisite phosphorylation
with a Hill function system and study the dynamic behavior of the fully
phosphorylated form ${\rm B}_n$.

The Hill equation is widely used in biochemical networks to model fast
signal response and complex binding processes. It was first introduced to
model the observed curves of ligand binding to the
receptor~\cite{hill1910possible}. The equation is a nonlinear function of
the ligand concentration, and the Hill exponent defines the degree of
cooperativity of ligand binding. Other complex models have been proposed
to describe different cooperativity of ligand
binding~\cite{adair1925hemoglobin,pauling1935oxygen,koshland1966comparison,monod1965nature}.
The sigmoidal curves generated from the Hill function system is similar
to the switchlike protein activity. Furthermore, the Hill equation requires
little prior knowledge of the binding mechanism and is much simpler than
the kinetic model~\eqref{eq:multi}. Therefore, this work investigates the
simple Hill function system for the ordered distributive enzyme-substrate
system~\eqref{sys:multi}:
 \begin{equation}\label{sys:hill}
{\rm B}_0 \xrightleftharpoons[\displaystyle k_d]{\displaystyle k_a\frac{[{\rm
A}]^\sigma}{k_m^\sigma+[{\rm A}]^\sigma}} {\rm B}_n,
 \end{equation}
where $k_a$ and $k_d$ are the forward and reverse reaction constant,
respectively, $k_m$ is the dissociation constant, and $\sigma$ is the Hill
exponent, a real number with range $\sigma \leq n$.

Since the Hill equation has been thoroughly studied and extensively used
in traditional deterministic models (differential equations), the unexplored
discrete stochastic representation gets more attention from researchers
with the presence of low population levels and molecular noise. Previous
studies have discovered that the sigmoidal behavior of the Hill function
dynamics may reduce to a linear function in the stochastic regime, especially
under the reaction-diffusion master equation
framework~\cite{chen2017stochastic}. This paper models the Hill
function system in the stochastic regime and restricts attention
to a homogeneous domain. Using the stochastic simulation algorithm
(SSA)~\cite{gillespie1976general}, the above Hill function
system~\eqref{sys:hill} is then governed by the forward and reverse reactions
with propensities:
 \begin{equation}
    a_f = k_a\frac{[{\rm A}]^\sigma}{k_m^\sigma+[{\rm A}]^\sigma}[{\rm B}_0];
    \quad a_r = k_d[{\rm B}_n].
 \end{equation}

Biologists are able to quantify species population at molecular levels
with improved experimental techniques~\cite{raj2009single}, which provides
good resources (observed empirical data) for validating stochastic models
and optimizing ``stochastic parameters''. Assuming the ordered distributive
mechanism~\eqref{sys:multi} is the ground truth for the multisite
phosphorylation process, the trajectories of $[{\rm B}_n]$ generated from
the stochastic enzyme-substrate system~\eqref{sys:multi} can be considered
empirical data, while the trajectories of $[{\rm B}_n]$ generated from the
stochastic Hill function system \eqref{sys:hill} are considered as simulated
data. This paper, instead of validating the stochastic results against
the deterministic results, will focus on optimizing the stochastic Hill
function model for the multisite phosphorylation process, meaning optimizing
the parameter vector $\theta= [k_a$, $k_d$, $\sigma$, $k_m]$ defining the
stochastic Hill function model.

The challenges of solving the inverse problem of parameter estimation for
modeling biological systems are manifold. One major challenge is large
biological networks with many unknown parameters that are hard or impossible
to measure in experiments. The nonlinear nature of the models involves a
nonconvex problem with multiple locally optimum points, and local
optimization methods may be trapped at local optimum points. Global
optimization methods can be very expensive in high dimensions, and global
optimality is often not guaranteed \cite{ashyraliyev2009systems}. Moreover,
most of the efforts in solving such problems thus far have been focused
on deterministic models, particularly estimating the parameters of models
formulated by nonlinear differential equations~\cite{mendes1998non}.
Parameter estimation in stochastic models is even more challenging as the
amount of empirical data must be large enough to obtain statistically valid
parameter estimates. Two well-known approaches for stochastic optimization
problems are stochastic approximation (SA) and response surface methodology
(RSM). The class of quasi-Newton methods for stochastic optimization
extends state-of-the-art numerical optimization methods (e.g., secant
updates, trust regions)~\cite{castle2012quasi}, can also be used for
deterministic global optimization with minor
variations~\cite{easterling2014parallel, amos2014algorithm}, and has been
successfully applied to various stochastic optimization problems, such as
cell cycle models~\cite{chen2017quasi}, bistable
models~\cite{chen2018parameter}, and biomechanics
problems~\cite{radcliffe2010results}. Other approaches such as Bayesian
inference~\cite{liu2017parameter, robert2013monte}, Kalman filter, and its
variants~\cite{liu2012state} have been applied to solve this problem as
well. Recently, machine learning techniques have been tailored by
sparsity-promoting methods to identify not only the parameters but also
the structure of both ordinary differential
equations~\cite{brunton2016discovering} and partial differential
equations~\cite{rudy2017data}.

Most parameter optimization methods only return a single best parameter
vector, regardless of the fact that there are many parameter vectors that
could generate similar system dynamics and characteristics. Take the
bistable switch in the cell cycle as an example~\cite{gerard2013minimal}.
Bistable phenomena occur when model parameters are inside the bistability
region. The entire bistability region or at least most parameter values
sampled in the region may be considered acceptable if the goal is to simply
model the bistable switch process, rather than to minimize the objective
function. This work focuses on finding an ``acceptable" parameter region,
where parameter vectors in the acceptable region are good alternatives to
the best parameter vector minimizing the objective function. In other
words, the system parameters are given by a region in parameter space
rather than a single point.

This work studies acceptable parameter regions of the stochastic Hill
function~\eqref{sys:hill} for the multisite phosphorylation
system~\eqref{sys:multi}. Section~\ref{sec:background} introduces the
chemical master equation for deriving the transition probability matrix
and the quasi-Newton algorithm used for stochastic optimization. In
Section~\ref{sec:objFunc}, three objective functions measuring different
features of the simulation-based empirical data are investigated.
Section~\ref{sec:rule} then presents the proposed $\alpha$-$\beta$-$\gamma$
rule for defining the acceptable parameter regions found by the quasi-Newton
stochastic optimization (QNSTOP) algorithm. Numerical results and detailed
analyses are given in Section~\ref{sec:results}.

\section{Background}~\label{sec:background}
This section first reviews the chemical master equation and the analytical
solution of a general biochemical system. Then, the essential steps of the
quasi-Newton stochastic optimization algorithm (QNSTOP) are summarized.

\subsection{Chemical Master Equation}
Consider a well-mixed system of $N$ distinct species and $M$ reaction
channels with $\hat N$ possible states. The chemical master
equation~\cite{gillespie1992rigorous} that describes the probabilistic
time evolution of the system dynamics is
 \begin{equation} \label{eq:cme}
\frac{d}{d t}{{P}}({X}; t) =
    {P}({X}; t) {C} ,
 \end{equation}

where ${X} = [{x_1}, {x_2}, \ldots, {x_{\hat N}}]$ is all possible state
vectors $x_i$ at any time $t$, $P({X}; t)$ represents the probabilities
of those state vectors at time $t$, and ${C}$ is the state reaction
matrix~\cite{munsky2006finite}, given by
 \begin{equation}\label{eq:A_matrix}
	{C}_{ij}=\left\{ \begin{array}{c c}
		-\sum_{\mu=1}^{M}a_{\mu}({x}_j), &  \text{for } i=j, \\
		a_{\mu}({x}_i),  & \text{for } i \text{ such that } {x}_j
		= {x}_i + v_{\mu},\\ 0,  & \text{otherwise},
	\end{array} \right.
 \end{equation}
where $v_{\mu}$ is the stoichiometric transition vector for reaction channel
$\mu$.

From equation \eqref{eq:cme}, the solution is
\begin{equation}\label{eq:solution}
	{P} = P_0e^{Ct}, 
\end{equation}
where $P_0$ is the initial probability of all the possible states, and 
the transition probability matrix can be calculated by 
\begin{equation}\label{eq:E_matrix}
	{\mathcal T} = e^{{C}\tau}, 
\end{equation}
where $\tau$ is the period of time the system has evolved from a previous time~\cite{munsky2006finite}.

\subsection{Quasi-Newton Stochastic Optimization Algorithm}
QNSTOP is a class of quasi-Newton methods developed for stochastic
optimization~\cite{castle2012quasi}, where the objective function $f(X)$
is a random variable, and $X$ is contained in a box $L \le X \le U$. The
essential steps of QNSTOP are summarized here. Further details on the
algorithm and implementation can be found in Ref.~\cite{amos2014algorithm}.

\begin{itemize}
\item In each iteration $k$, construct a quadratic model
    \begin{equation*}
        \widehat{m}_k(X-X_k) =	{\hat f}_k + {\hat g}_k^T \left( X-X_k \right) + \frac{1}{2}	\left(X-X_k\right)^T {\hat H}_k \left(X-X_k\right) 
    \end{equation*}
centered at $X_k$, where ${\hat g}_k$ is the gradient vector and
${\hat H}_k$ is the Hessian matrix.
Note that ${\hat f}_k$ is generally not ${f}{(X_k)}$, which is stochastic.

\item QNSTOP uses an ellipsoidal region centered
at the current iterate $X_k \in \real^n$ with radius $\tau_k$,
\begin{equation*}
    E_k(\tau_k) = \left\{ X \in \real^n: \left( X-X_k \right)^T W_k
\left(X-X_k\right) \le \tau_k^2 \right\},
\end{equation*}
where $W_k$ is a symmetric, positive definite scaling matrix, satisfying
$W_k \in W_\gamma$, 
\begin{equation*}
    {W_\gamma = \bigl\{ W \in \real^{n \times n} : W = W^T,  \det(W) = 1,
\gamma^{-1} I_n \preceq W \preceq \gamma I_n \bigr\}}
\end{equation*}
for some $\gamma \ge 1$, where $I_n$ is the $n \times n$ identity matrix,
and $A \preceq B$ means $B-A$ is positive semidefinite.
The elements of the set ${\bf {\rm W}}_\gamma$ are valid scaling matrices
that control the shape of the ellipsoidal design regions with eccentricity
constrained by $\gamma$. 

\item Then QNSTOP estimates the gradient based on a set of $N$ uniformly
sampled design sites $\{X_{k1}$, $\ldots$, $X_{kN}\} \subset E_k(\tau_k)
\cap \Theta$ ($\Theta=[L,U]$, which is the feasible set of parameters
defined initially.) For the Hessian matrix, QNSTOP uses either a variation
of the SR1 (symmetric, rank one) quasi-Newton update (stochastic $f$) or
the unconstrained BFGS quasi-Newton update (global optimization of
deterministic $f$).

\item For the next iteration, by utilizing an ellipsoidal trust region
concentric with the design region for controlling step length, $X_{k+1}$
is updated as
 \begin{equation*}
X_{k+1} = \left( X_k - \left[{\hat H}_k + \mu_k W_k\right]^{-1} \hat g_k
\right)_{\Theta},
 \end{equation*}
where $\mu_k$ is the Lagrange multiplier of a trust region subproblem, and
$\left(\cdot\right)_{\Theta}$ denotes projection onto the feasible set
$\Theta=[L,U]$.

\item Finally, the experimental design region $E_k(\tau_k)$ is updated to
approximate a confidence set by updating the scaling matrix $W_k$. The
updated scaling matrix is given by
\begin{equation*}
   W_{k+1} = \left(\hat H_k + \mu_k W_k\right)^T V_k^{-1} \left(\hat H_k
+ \mu_k W_k\right), 
\end{equation*}
where $V_k$ is the covariance matrix of $\nabla \widehat{m}_k(X_{k+1}-X_k)$.
For numerical stability, $W_{k+1}$ is constrained (by modifying its
eigenvalues) to satisfy the constraints $\gamma^{-1} I_n \preceq W_{k+1}
\preceq \gamma I_n$ and $\det(W_{k+1}) = 1$, so $W_\gamma \ni W_{k+1}$.

\end{itemize}


\section{Objective Functions}~\label{sec:objFunc}
This section presents three different objective functions characterizing
different aspects of the population trajectory of ${\rm B}_n$. In particular,
we propose a general simulation-based objective function that can be applied
to large biochemical networks.

\subsection{Minimum distance area}
In the enzyme-substrate system~\eqref{sys:multi}, suppose $D = [\hat x_1$,
$\hat x_2$, $\ldots$, $\hat x_m]$ is a sequence of the molecular population
of ${\rm B}_n$ collected from stochastic simulation results after every time
$\tau$ (denoted as $[t_1$, $t_2$, $\ldots$, $t_m]$), where $m$ is the data
size. This subsection will consider the population difference of ${\rm
B}_n$ between the empirical data (from multisite phosphorylation) and
simulation results (from stochastic Hill function) over time, called the
distance area. Given the empirical data and the simulated data as two
vectors, the $p$-norm is one traditional way to measure the vectors'
difference. The problem is stochastic and the time series data can be
quite noisy, and outliers have more influence for $p>1$, hence the 1-norm
is used to measure the distance. Define the distance area as the objective
function,
 \begin{equation}
	f_d(\theta) = \int |p(t)-q(t)|dt \approx \sum_{i=1}^{m} |\hat
 x_i-\hat y_i| \tau,
\end{equation} where $\theta$ is the vector of model parameter values,
$p(t)$ and $q(t)$ are the trajectory functions of empirical and simulated
populations in continuous domains. For discrete time series data, $\hat
x_i$ and $\hat y_i$ represent the population of ${\rm B}_n$ from empirical
and simulated data, respectively. The stochastic optimization problem to
be solved is
 \begin{equation}
	\min_{\theta \in \Theta} f_d(\theta),
 \end{equation}
where $\Theta\subset {\rm I\!R}^n$ defines the feasible set (allowable
values for the model parameter vector $\theta$).

\subsection{Maximum log-Likelihood}
The minimum distance area, similar to other traditional optimization
methods, builds objective functions based on `mean' measurements from
stochastic simulations, hence cannot reflect the intrinsic noise in
stochastic models. To capture the stochastic fluctuations, measure the
transition probability that a system jumps from one state to the next state
after a certain time step. The likelihood function of time series data
can then be factorized into the product of transition probabilities. For
convenience, the logarithm of the likelihood, which changes a product to
a sum, is used. The logarithm of the likelihood of the observed data $D$
is
 \begin{equation}
	\log \mathcal{L} ({\theta} | {D}) = \log  \bigg( \prod_{i=2}^{m}
	{{\mathcal T}}_{x_{i-1},x_{i}} \bigg) = \sum_{i=2}^{m} \log
	{{\mathcal T}}_{x_{i-1},x_{i}}
 \end{equation}
where $\theta \in {\rm I\!R^n}$ is the vector of model parameter values
and ${\mathcal T}$ is the transition probability matrix. Specifically,
${\mathcal T}_{x_{i-1},x_i}$ is the transition probability that the system
changes from state $x_{i-1}$ to state $x_i$. Note that we take the logarithm
of the likelihood because the transition probability matrix is usually
very close to zero. The log-Likelihood function expresses the probabilities
of the observed empirical data for different values of parameter vector
$\theta$. A larger value of log-likelihood indicates a better fit to the
empirical data.

Using the maximum log-likelihood, the objective function is 
 \begin{equation}
	f_l(\theta) = -\log \mathcal{L} ( {\theta} | {D}),
 \end{equation}
and the stochastic optimization problem to be solved is
 \begin{equation}
	\min_{\theta \in \Theta} f_l(\theta),
 \end{equation}
where $\Theta$ is a set in ${\rm I\!R^n}$ defining the feasible set
(allowable values for the model parameter vector $\theta$).

When a system has a finite number of states, then we can calculate $\mathcal
T$ directly from Eq.~\eqref{eq:E_matrix}. When a system is small and has
an infinite number of states, the finite state projection (FSP)
method~\cite{munsky2006finite} projects the infinite state vector ${X}$
to a finite state vector, approximating the CME solution with an error
$\epsilon$. Accordingly, ${C}$ and ${\mathcal T}$ are approximated by
${\hat C}$ and ${ \hat {\mathcal T}}$, respectively. Fox et
al.~\cite{fox2016finite} proved that the FSP-derived likelihood converges
monotonically to the exact likelihood value.

\subsection{Approximate maximum log-likelihood}
In the above maximum log-likelihood method, the approximation of the
transition matrix ${\mathcal T}$ is only tractable for small systems. It
is difficult or nearly impossible to solve the CME for large systems. To
overcome this limitation, this subsection proposes an approximate maximum
log-likelihood method, a general use objective function that is applicable
to large complex biochemical networks.

The transition probability of system state going from $x_{i-1}$ to $x_i$ is
 \begin{equation}
	{\mathcal T}_{x_{i-1},x_i} ={\rm Pr}(x_i | x_{i-1}).
 \end{equation}
Thus, the logarithm of the likelihood of the empirical data $D$ is 
 \begin{equation}
	\log \mathcal{L} ({\theta} | {D}) =
	\log  \bigg( \prod_{i=2}^{m} {{\rm Pr}(x_i | x_{i-1})} \bigg) =
	\sum_{i=2}^{m} \log  {\rm Pr}(x_i | x_{i-1}),
 \end{equation}
where ${\rm Pr}(x_i | x_{i-1})$ can be approximated using simulation data
(An example will be shown later in~\eqref{eq:p_example}). The objective
function of approximate maximum log-likelihood is
 \begin{equation}
	f_p(\theta) = -\log \mathcal{L} ( {\theta} | {D}).
 \end{equation}
The stochastic optimization problem to be solved is
 \begin{equation}
	\min_{\theta \in \Theta} f_p(\theta),
 \end{equation}
where $\Theta$ is a set in ${\rm I\!R^n}$ defining the feasible set.

Algorithm 1 summarizes the essential steps of the approximate maximum
log-likelihood method. In Line 6, by simulating the system $q$ times from
time $t_{i-1}$ to $t_i$ with initial system state $[{\rm B}_n]= x_{i-1}$,
we get a list of simulation results for $[{\rm B}_n]$ at time $t_{i}$,
represented as $S_{i}=[s_1$, $s_2$, $\ldots$, $s_q]$. ${\rm Pr}(x_i |
x_{i-1})$ can be approximated from the distribution of $S_i$. For example,
assuming $x_i$ (sampled in $S_{i}$) follows a normal distribution
 \begin{equation}
x_i \sim N(\mu, \sigma^2),
 \end{equation}
where $\mu$ and $\sigma^2$ are the mean and variance of $x_i$, we can
approximate the probability as
 \begin{equation}
\label{eq:p_example}
	{\rm Pr}(x_i | x_{i-1}) \approx {\rm Pr}(x_i-0.5< \mu + \sigma z_i
	<x_i+0.5), \quad z_i = \frac{x_i-\mu}{\sigma} \sim N(0,1).
 \end{equation}
Note that the empirical data could be multiple species' population
trajectories, then $x_i$ in ${\rm Pr}(x_i | x_{i-1})$ becomes a vector
referring to multiple species' population. In Line 4, if the simulation
results do not include $[{\rm B}_n]=\hat x_i$, then choose one that is the
closest to $\hat x_i$.

\begin{algorithm}[htb]
  \caption{Approximate maximum log-likelihood}
  \begin{algorithmic}[1]
    \INPUT Empirical data $D=[\hat x_1$, $\hat x_2$, $\ldots$, $\hat x_m]$
    represents the population trajectory of species $[{\rm B}_n]$
    \OUTPUT $-\sum_{i=2}^{m} \log {\rm Pr}(\hat x_i | \hat x_{i-1})$ 
    \STATE \textbf{Initialization} $i=2$, system state $\hat x_1$.
    \WHILE{$i \leq m$}
      \IF{$i=2$} {go to line 6}
      \ELSE {~initialize the system with the simulation result where
      $[{\rm B}_n]=\hat x_{i-1}$ at time $t_{i-1}$}
      \ENDIF
      \STATE Simulate the system $q$ times from time $t_{i-1}$ to $t_{i}$
generating a list of simulation results for ${\rm B}_n$ at time $t_{i}$,
denoted as $S_{i}=[s_1$, $s_2$, $\ldots$, $s_q]$.
      \STATE Construct ${\rm Pr}(\hat x_i|\hat x_{i-1})$ based on the
distribution of $S_{i}$
      \STATE $i = i + 1$
    \ENDWHILE
  \end{algorithmic}
\end{algorithm}

\section{Acceptable Parameter Region}~\label{sec:rule}
As mentioned before, for most systems, especially those with multidimensional
parameters, there are possibly many combinations of parameter values
producing similar system dynamics and behaviours. This section introduces
the $\alpha$-$\beta$-$\gamma$ rule to find the acceptable parameter regions
of the stochastic Hill function. Parameter values sampled in an acceptable
region should have similar system results compared to the best parameter
values. In particular, we applied the $\alpha$-$\beta$-$\gamma$ rule to
QNSTOP, which has been used to find the best parameter values of several
stochastic problems.

\subsection{$\alpha$-$\beta$-$\gamma$ Rule} %
Based on the fact that QNSTOP creates an ellipsoidal design region at each
iteration, we can utilize this ellipsoid to define the acceptable parameter
region. For an ellipsoid $E$, define the objective function values as
$f(E)=\{f(x) \mid x \in E\}$. To accept a parameter region, intuitively,
most parameters sampled from the region should have relatively small
objective values, and close to the minimum objective function value of the
ellipsoidal region, written as $\min f(E)$. Based on scrutinizing all
possible distributions of $f(E)$, define a stable ellipsoidal region $E$
as follows:

\begin{definition}{Assume $f(\theta)\ge 1$ always.
$E$ is a min-stable region} if
\begin{equation*}
{\rm Pr}\big [ f(\theta) \le (1+\alpha) \min f(E) \big ] \ge \beta, \quad \theta \in E, \ \alpha \in (0, \infty), \ \beta \in (0, 1].
\end{equation*}
\end{definition}

In the above definition, $\alpha$ measures how close the objective function
values are to the minimum value $\min f(E)$, $\beta$ controls the percentage
of parameter values that generate close minimum objective function values.
$\alpha$ and $\beta$ can be assigned with different values depending on
the problem. For any parameter region $E$, if $\alpha$ is fixed, define
the percentage of points with objective function values no larger than
$(1+\alpha)\min f(E)$ as the \textbf{region stability}, which is the value
of ${\rm Pr}\big [ f(\theta) \le (1+\alpha) \min f(E) \big ]$.

As QNSTOP designs a specific ellipsoidal region at each iteration, the minimum
objective function values found may vary dramatically between iterations.
We don't want to accept the ellipsoidal region $E_1$ of the first iteration
even if $E_1$ is min-stable, because $\min f(E_1)$ is usually much larger
than the minimum objective value $f_{min}$ found over all iterations and
all starting points (if QNSTOP is run with multiple starting points). Thus,
to ensure that the parameter region is globally min-stable, we need to
choose those min-stable regions whose local minimum objective function
values are close to the lowest minimum found so far. The acceptable
parameter region is defined as a union of min-stable ellipsoids with local
minimum objective function values close to $f_{min}$:

\begin{definition}{An acceptable region} is $\displaystyle
R=\bigcup_{k \in {\mathcal B}} E_k$ where
\begin{equation*}
{\mathcal B} = \{ k \ | \ E_k \ \text{is min-stable and} \ \min f(E_k) \le (1+\gamma)f_{min} \big \}, \quad \gamma \in [0, \infty).
\end{equation*}
\end{definition}

In the above definition, $k$ is the iteration number, $\gamma$ controls
how close $\min f(E_k)$ is to the lowest minimum found over all iterations,
and $\gamma$ may vary according to the problem. Choosing values for $\alpha$,
$\beta$, $\gamma$ is referred to as the $\alpha\hbox{-}\beta\hbox{-}\gamma$
rule. Note that if optimization algorithms use multiple starting points,
then $k$ indexes iterations over all starting points, and $f_{min}$ is the
minimum found over all starting points.

\subsection{Analysis}
\begin{figure}[!htb]
\centering
\begin{subfigure}{0.45\textwidth}
\centering
\includegraphics[width=1\textwidth]{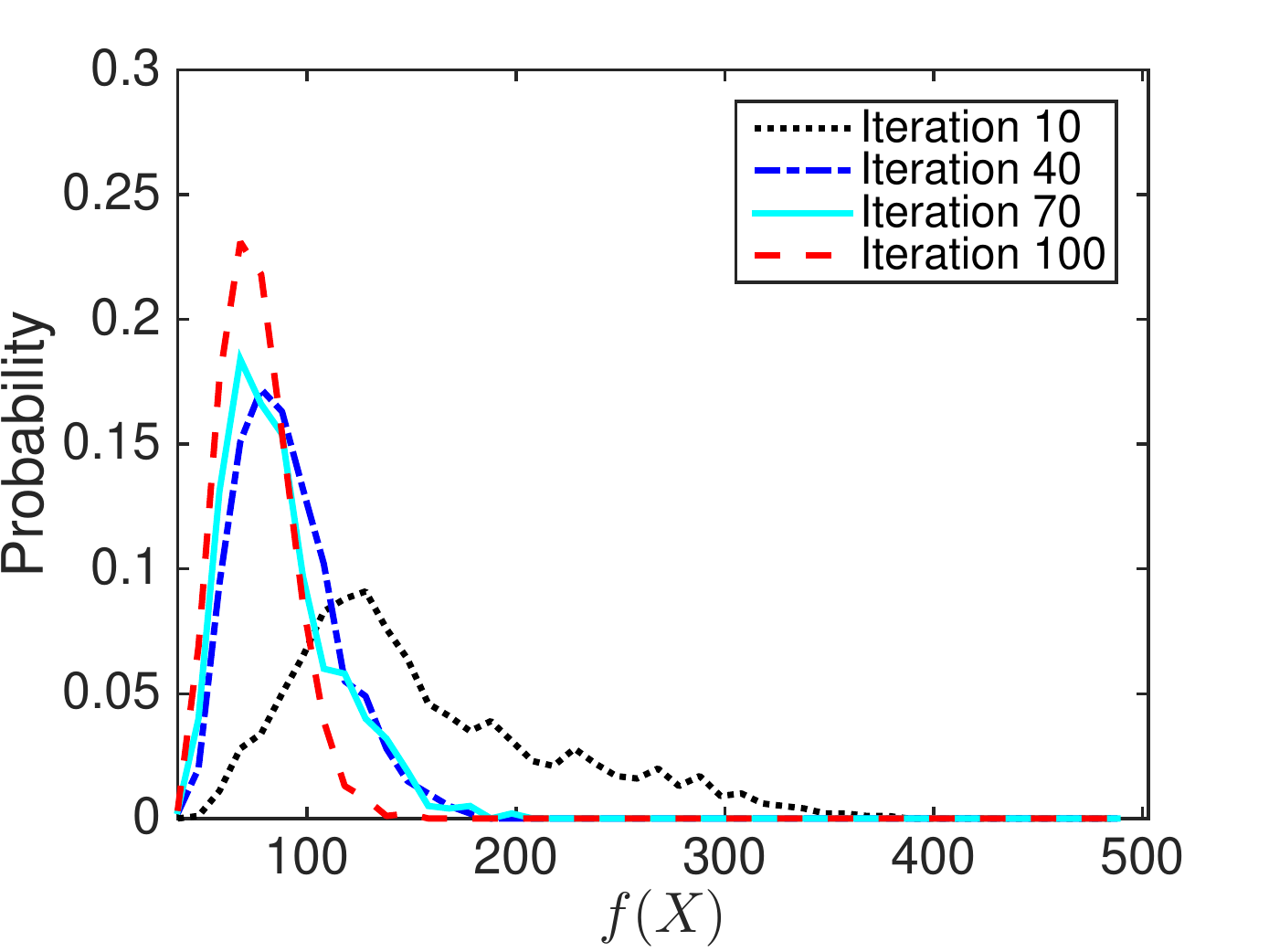}
\caption{Minimum distance area}
\end{subfigure}
\hspace{2em}
\begin{subfigure}{0.45\textwidth}
\centering
\includegraphics[width=1\textwidth]{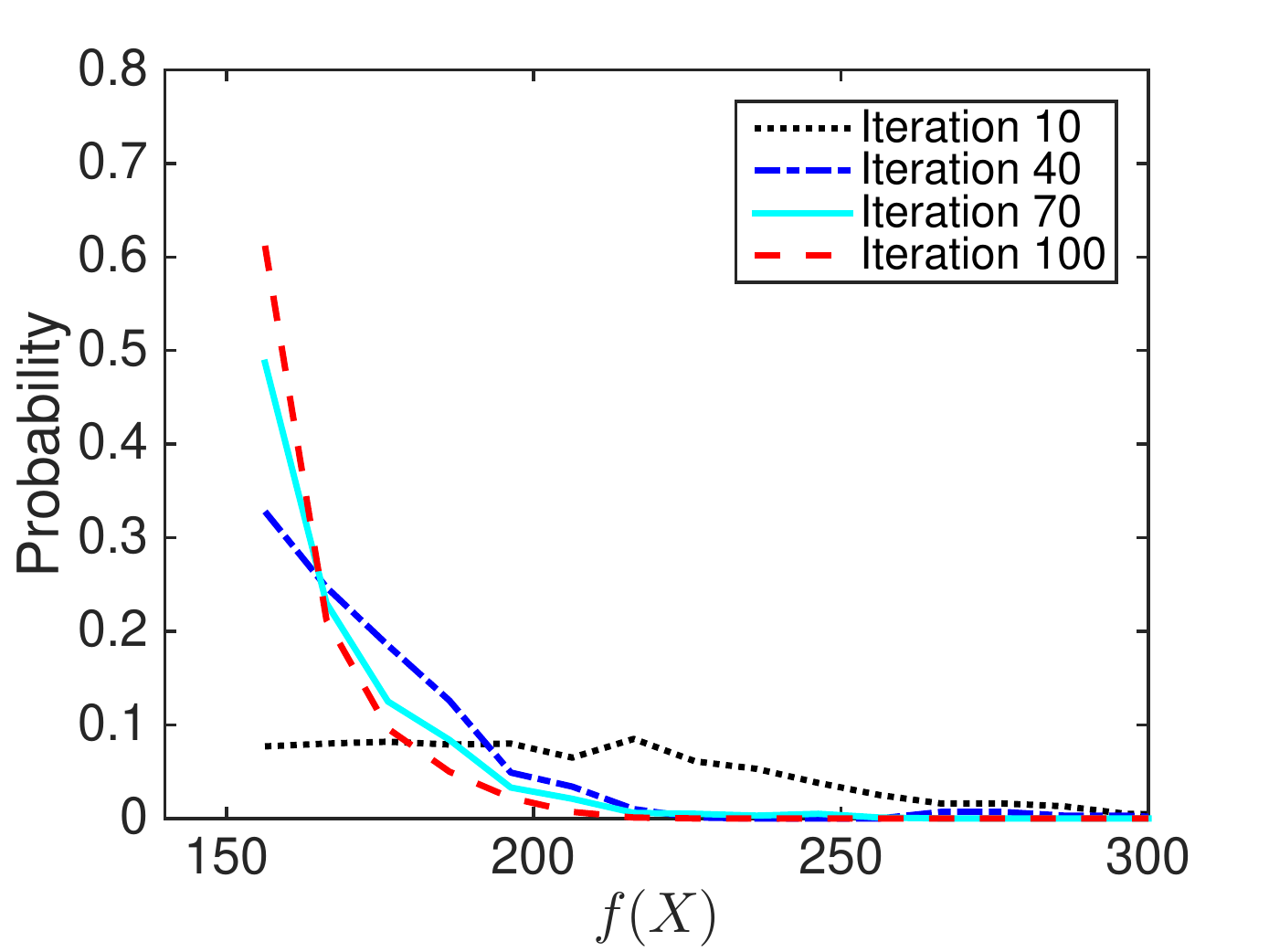}
\caption{Maximum log-likelihood}
\end{subfigure}
\begin{subfigure}{0.45\textwidth}
\centering
\includegraphics[width=1\textwidth]{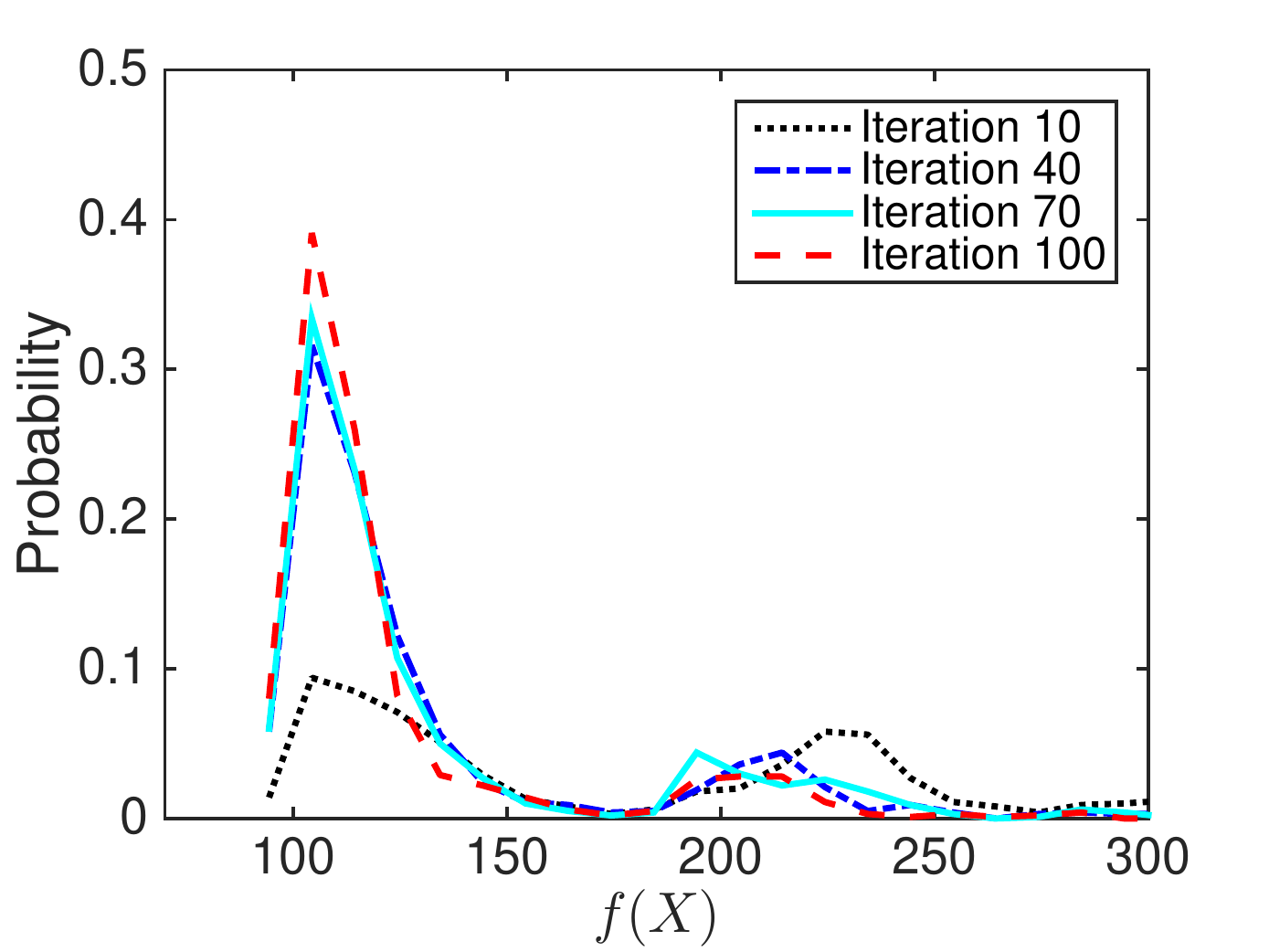}
\caption{Approximate maximum log-likelihood}
\end{subfigure}
\caption{{Distributions of objective function values from three methods (minimum distance area, maximum log-likelihood, and approximate maximum log-likelihood) based on 1000 sampled points inside the ellipsoidal regions for iterations 10, 40, 70, and 100.}}
\label{fig:f-distribution}
\end{figure}

\begin{figure}[!htb]
\centering
\includegraphics[width=0.7\textwidth]{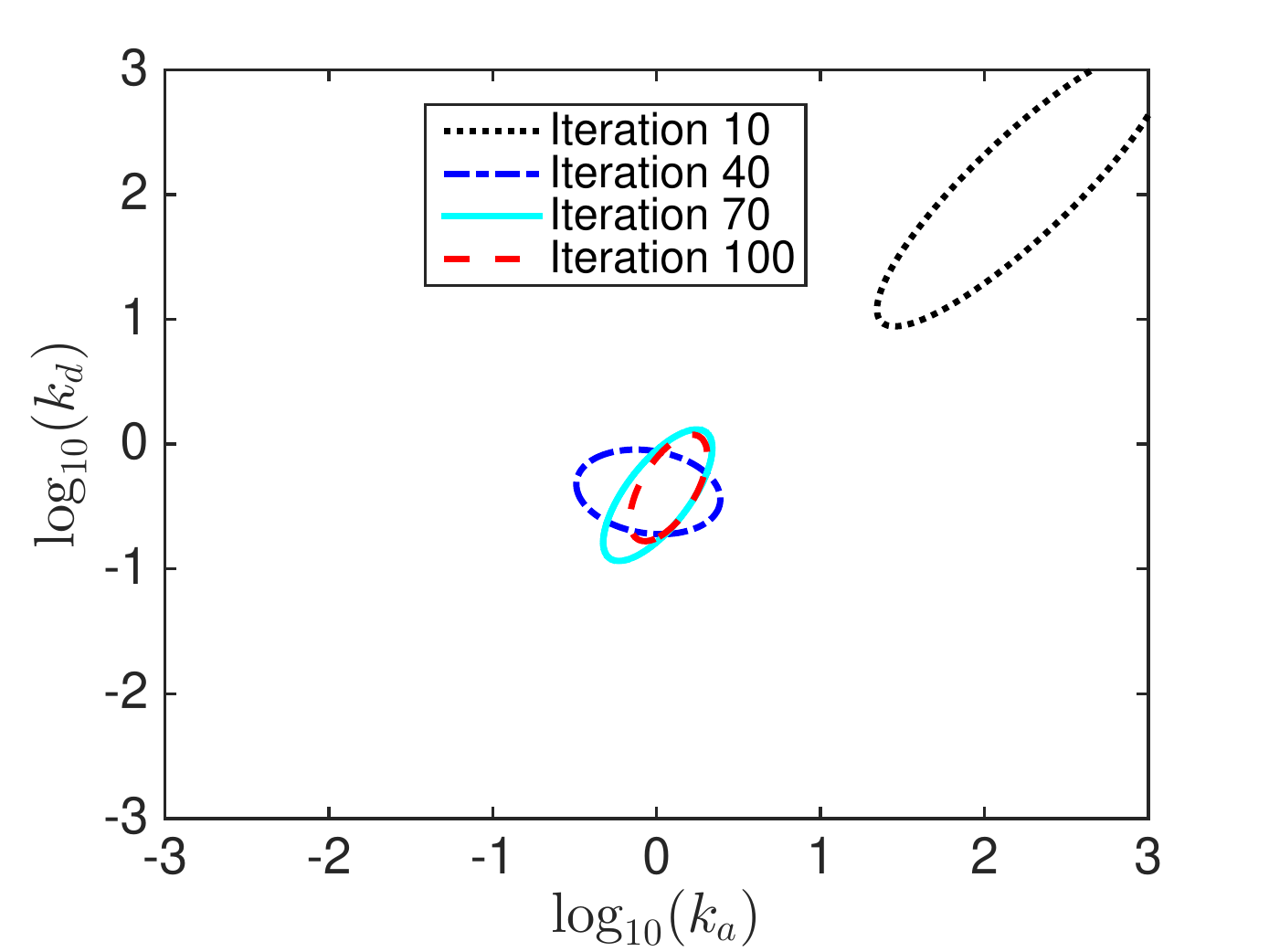}
\caption{QNSTOP ellipsoids at iterations 10, 40, 70, and 100 from the maximum log-likelihood method.}
\label{fig:f-iteration}
\end{figure}

Values for $\alpha$, $\beta$, and $\gamma$ are derived from analyzing the
objective function values in parameter regions.
Fig.~\ref{fig:f-distribution} shows the distributions of the three objective
function values over several iterations. For iteration 10, $f(X)$ of
maximum log-likelihood method is almost a uniform distribution, with values
spread out over the domain. As the iteration continues, more objective
function values are getting closer to $f_{min}$, the minimum value found
so far. The distribution of $f(X)$ gradually forms a peak around $f_{min}$.
While these features hold for all three methods, the maximum log-likelihood
has the highest percentage of the small objective values. Based on the
distributions, the values of $\alpha$ for the three methods are chosen as
0.5 (minimum distance area), 0.2 (maximum log-likelihood), 0.3 (approximate
maximum log-likelihood), respectively. Fig.~\ref{fig:f-iteration}
illustrates how the ellipsoidal regions of parameters $k_a$, $k_d$ shrink
over iterations.

\begin{figure}[!htb]
\centering
\includegraphics[width=0.7\textwidth]{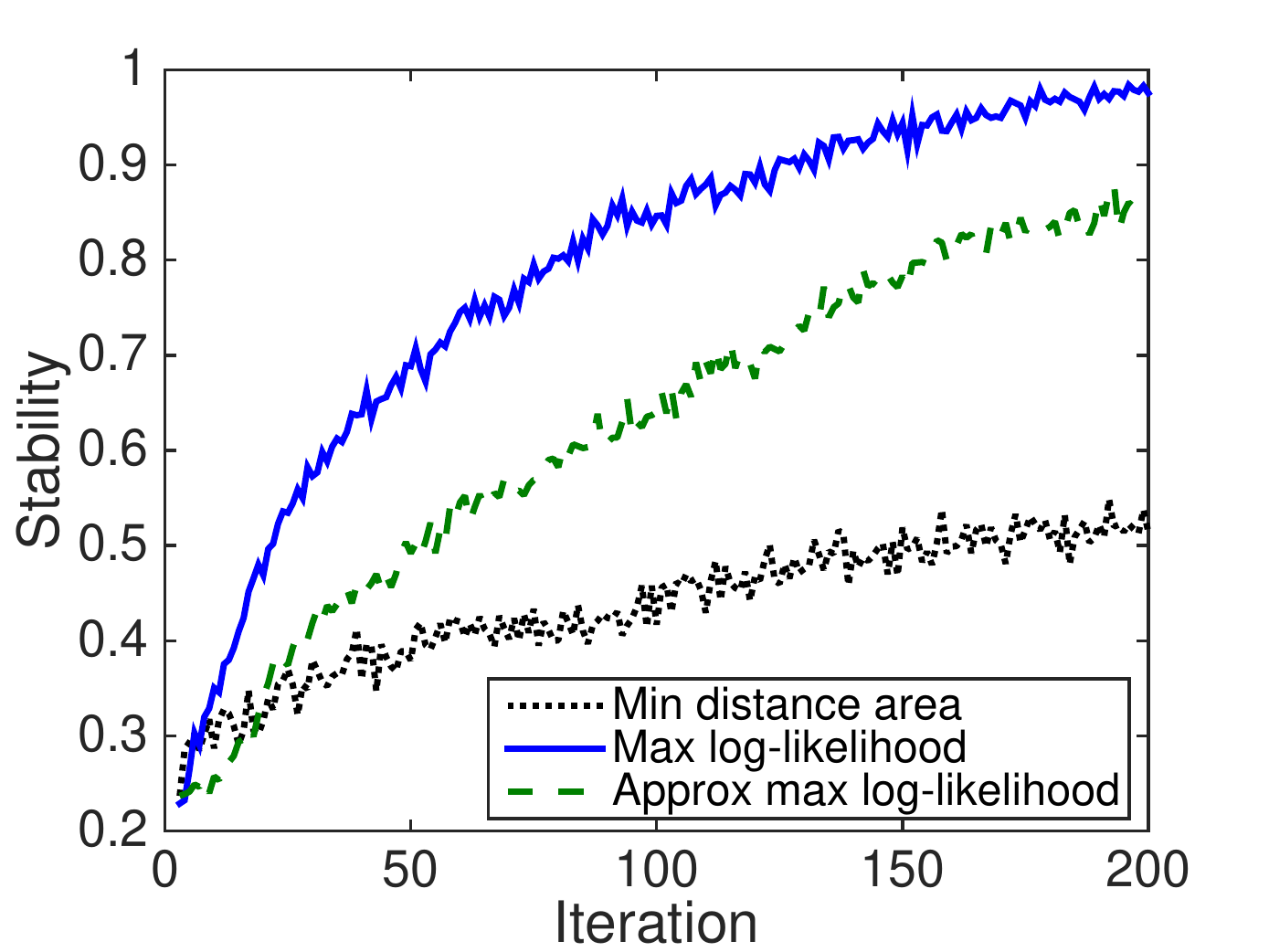}
\caption{Average region stability from iteration 1 to 200 based on 100 starting points from minimum distance area ($\alpha=0.5$), maximum log-likelihood ($\alpha=0.2$), and approximate maximum log-likelihood ($\alpha=0.3$).}
\label{fig:f-stability}
\end{figure}

Figure~\ref{fig:f-stability} shows that for all three methods, the average
region stability (defined before) over 100 starting points increases with
iteration number. The maximum log-likelihood has the highest region stability
compared to the other two objective functions. For the maximum
log-likelihood method, $\alpha=\gamma=0.2$ and $\beta=0.8$ based on the
region stability. In this way, at least 80\% of the sampled points from
the acceptable region have objective function values within 20\% relative
error of the minimum value, called the 80\%-20\% rule.
For the other two methods, $\alpha=\gamma=0.3$, $\beta=0.7$ for
approximate maximum log-likelihood, and $\alpha=\beta=\gamma=0.5$ for
minimum distance area.

\section{Results}~\label{sec:results}
This section first discusses the experimental setup including the empirical
data and input parameters. Then the optimization results are divided into
two parts. The first part demonstrates the result of optimizing two
parameters, $k_a$ and $k_d$; the second part compares the three
objective functions and studies the full parameter vector $\theta$.

\subsection{Experimental setup}
\textbf{Empirical data.} This work assumes the ordered distributive multisite
phosphorylation system~\eqref{sys:multi} as the ground truth. In particular,
consider the system size $n=4$, the population level of enzyme $[{\rm
A}]=1000$, and the reaction constants $f_i=0.0025$, $b_i=1$ for $i=1$,
$2$, $3$, $4$. The initial condition is $[{\rm B}_0]=100$, $[{\rm B}_i] =
0$ for $0<i\le n$. Use the stochastic simulation algorithm to simulate
the system and sample one population trajectory of ${\rm B}_n$ with a time
step $\tau$ $(t_1=\tau$, $t_2=2\tau$, $\ldots$, $t_m=m\tau)$ as a single
set $D$ of empirical data. Since both systems~\eqref{sys:multi}
and~\eqref{sys:hill} stabilize at a steady state after a certain time
(transition period), if the empirical data $D$ contains more steady state
information than transition period, then the system parameters will be
optimized in a way that minimizes the difference of steady states but
overlooks the transition dynamics before the system stabilizes, and vice
versa. Therefore, the empirical data points are sampled equally covering
both the transition and stable periods.

The stochastic Hill function system~\eqref{sys:hill} has an initial condition
$[{\rm B}_0]=100$ and $[{\rm B}_n]=0$, thus there are at most 101 system
states. Table~\ref{tab:para} lists the bounds for each parameter in the
system.
\begin{table}[hbt!]
	\centering
	\renewcommand{\arraystretch}{1.3}
	\caption{Parameter boundary in the stochastic Hill function system.}
	\label{tab:para}
	\begin{tabular}{ c  c  c  c  c }
		\toprule
		Parameter & $k_a$ & $k_d$ & $k_m$ & $\sigma$ \\ 
		\midrule
		$[L,U]$ & $[0.001, 1000]$ & $[0.001, 1000]$ & $[0.001, 10^6]$ & $[0.001, 10]$ \\
		$\log_{10} ([L,U])$ & $[-3, 3]$ & $[-3, 3]$ & $[-3, 6]$ & $[-3, 1]$ \\ 
		\bottomrule
	\end{tabular}
\end{table}

\begin{figure}[!htb]
\centering
\includegraphics[width=0.7\textwidth]{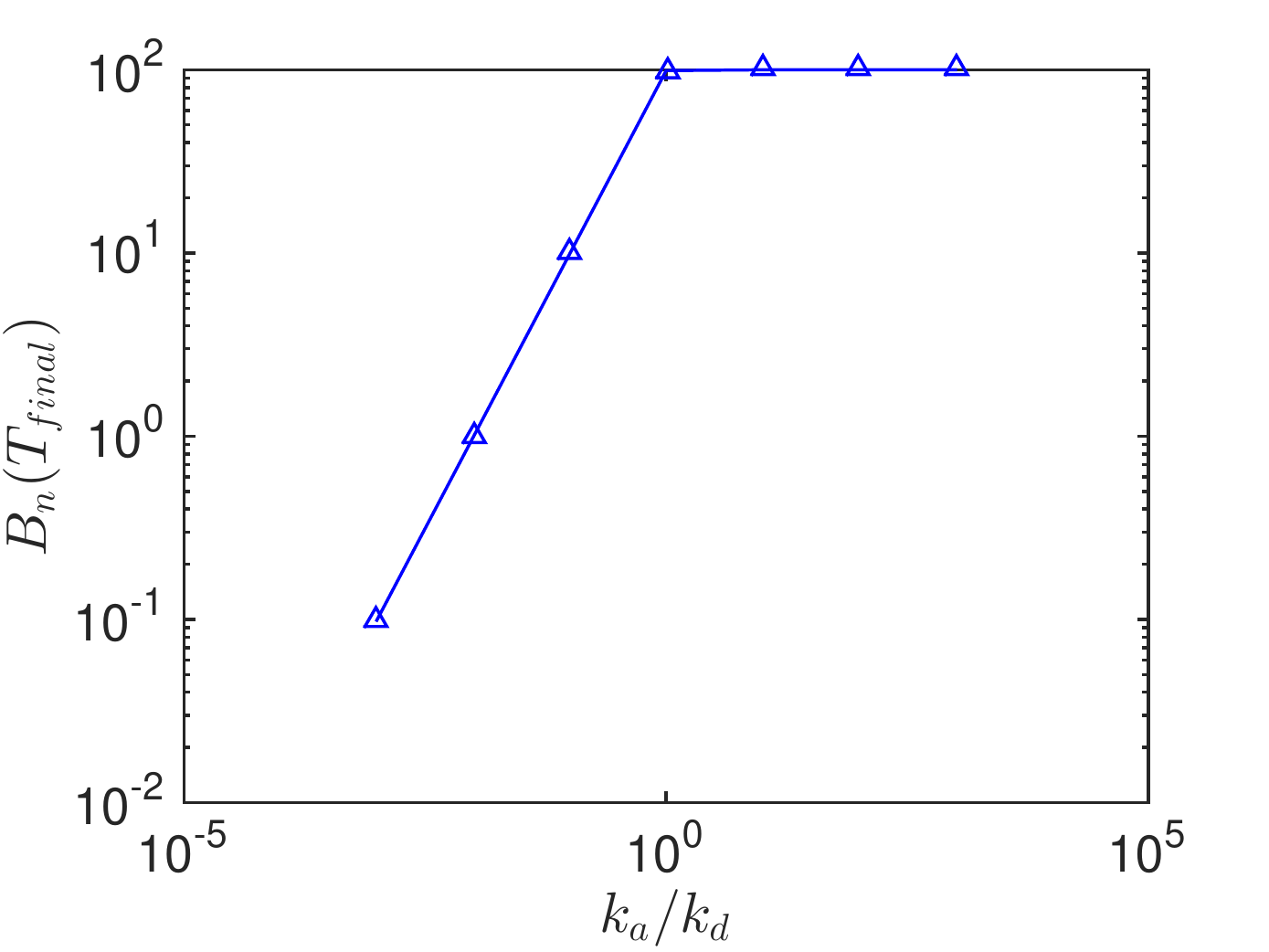}
\caption{{The population of ${\rm B}_n$ at stable state under the stochastic Hill function system with respect to different $k_a/k_d$ values, where $k_d=1$.}}
\label{fig:ks-kd}
\end{figure}

Note that parameters $k_a$ and $k_d$ (rate constants of association and
dissociation) control the system's time scale. For fixed parameters $k_m$
and $\sigma$, while parameter region $[0.001, 1]$ for $k_a$ and $k_d$
occupies 0.0001\% of the entire search box, the system time scale varies
by three orders of magnitude. In Fig.~\ref{fig:ks-kd}, the final population
of ${\rm B}_n$ initially grows linearly (in logarithm) with $k_a/k_d$ and
then levels off around 100. When $k_a>1$, $k_d=1$, the objective function
value does not change much because all ${\rm B}_0$ molecules are phosphorylated
at the stable state. Thus, the decimal region $[L, 1]$ ($L<1$ is the lower
bound), while sensitive, is minimized or overlooked when the upper bound
$U$ is much larger than one ($U \gg 1$). This \textbf{decimal parameter
sensitivity lost} phenomenon  can affect the optimization performance,
especially for systems that are sensitive to the $[0, 1]$ domain. To solve
this problem, simply use the logarithm of the parameters. Thus the bound
for $\log_{10}(k_a)$ and $\log_{10}(k_d)$ is $[-3, 3]$, shown in
Table~\ref{tab:para}. For numerical stability, QNSTOP scales the search
box $[L,U]$ to the unit cube.

\subsection{Two parameter case}
This subsection studies the two parameter case, in which $k_a$ and $k_d$
are unknown while $\sigma=2$ and $k_m=100$ in the stochastic Hill function
system. The QNSTOP settings are: total iterations = 100, sample point
number $N=10$, initial ellipsoid radius $\hbox{TAU}=0.85$ (one-tenth of
the $[L,U]$ box diameter), ellipsoid radius $\tau_k$ decay factor
$\hbox{GAIN}=35$, $\hbox{MODE}=\hbox{`G'}$ (global optimization mode). There are $m=50$ empirical
data points collected at time step $\tau=0.2$ from one SSA simulated
trajectory of the ${\rm B}_n$ population. The $\alpha$-$\beta$-$\gamma$
acceptable region is defined by $\alpha=0.2$, $\beta=0.8$, $\gamma=0.2$,
which indicates that 80\% of the sampled values should have objective
function values within 20\% relative error of the local minimum, and the
local minimum is within 20\% relative error of the lowest minimum found
over all starting points and iterations.

\begin{figure}[!htb]
\centering
\includegraphics[width=0.7\textwidth]{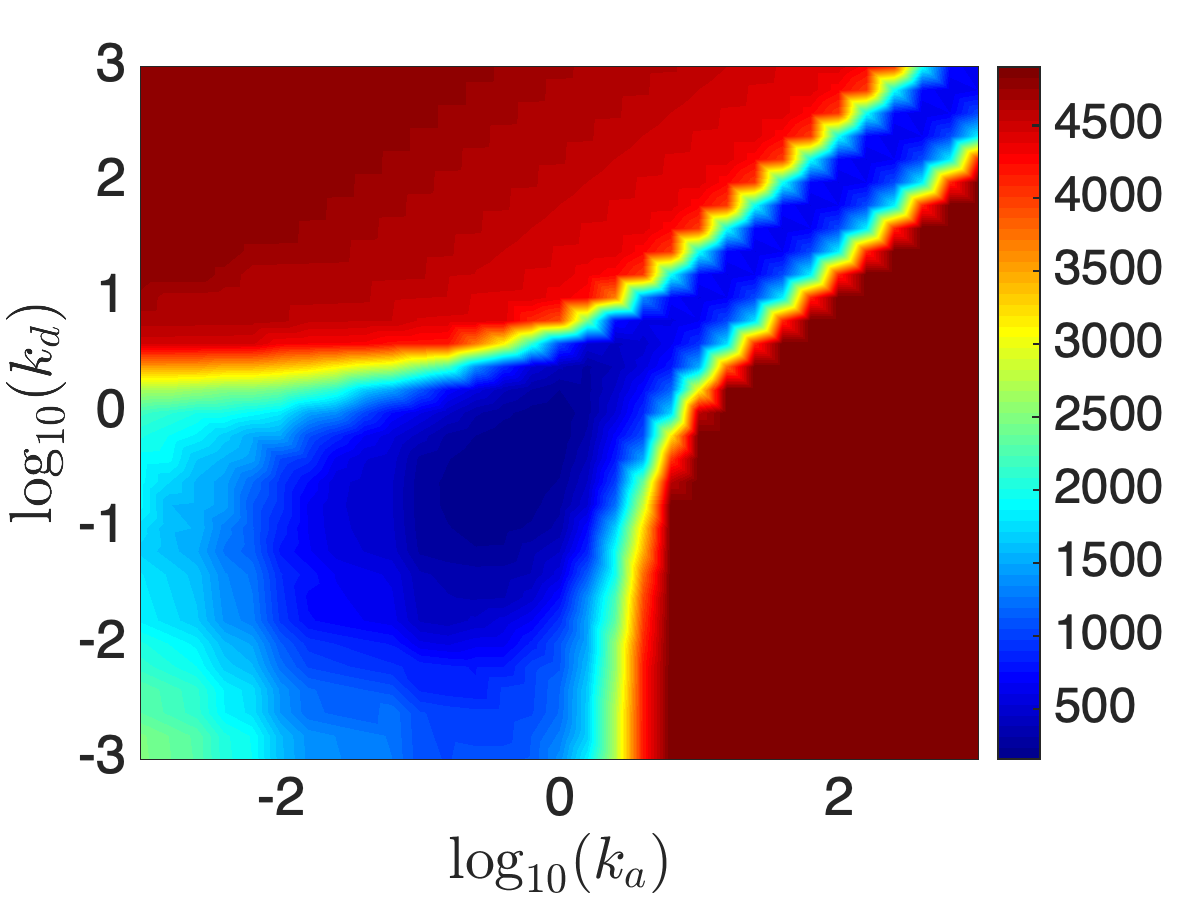}
\caption{{Exhaustive search of objective function values from the maximum log-likelihood method over domain $(\log_{10}(k_a), \log_{10}(k_d)) \in [-3, 3]^2$.}}
\label{fig:brutal}
\end{figure}

Fig.~\ref{fig:brutal} shows the objective function values of maximum
log-likelihood ($f_l$) over the entire $k_a$, $k_d$ domain. In the center
of the graph, the dark blue region ($\log_{10}(k_a)\in [-1, 0]$,
$\log_{10}(k_d) \in [-2, 0]$) has the minimum objective function value,
$f_l \approx 500$. Any ($\log_{10}(k_a)$, $\log_{10}(k_d)$) pairs sampled
in the dark blue region are acceptable values for the stochastic Hill
function system~\eqref{sys:hill}.

\begin{figure}[!htp]
\centering
\begin{subfigure}[]{0.45\textwidth}
\centering
\includegraphics[width=1\textwidth]{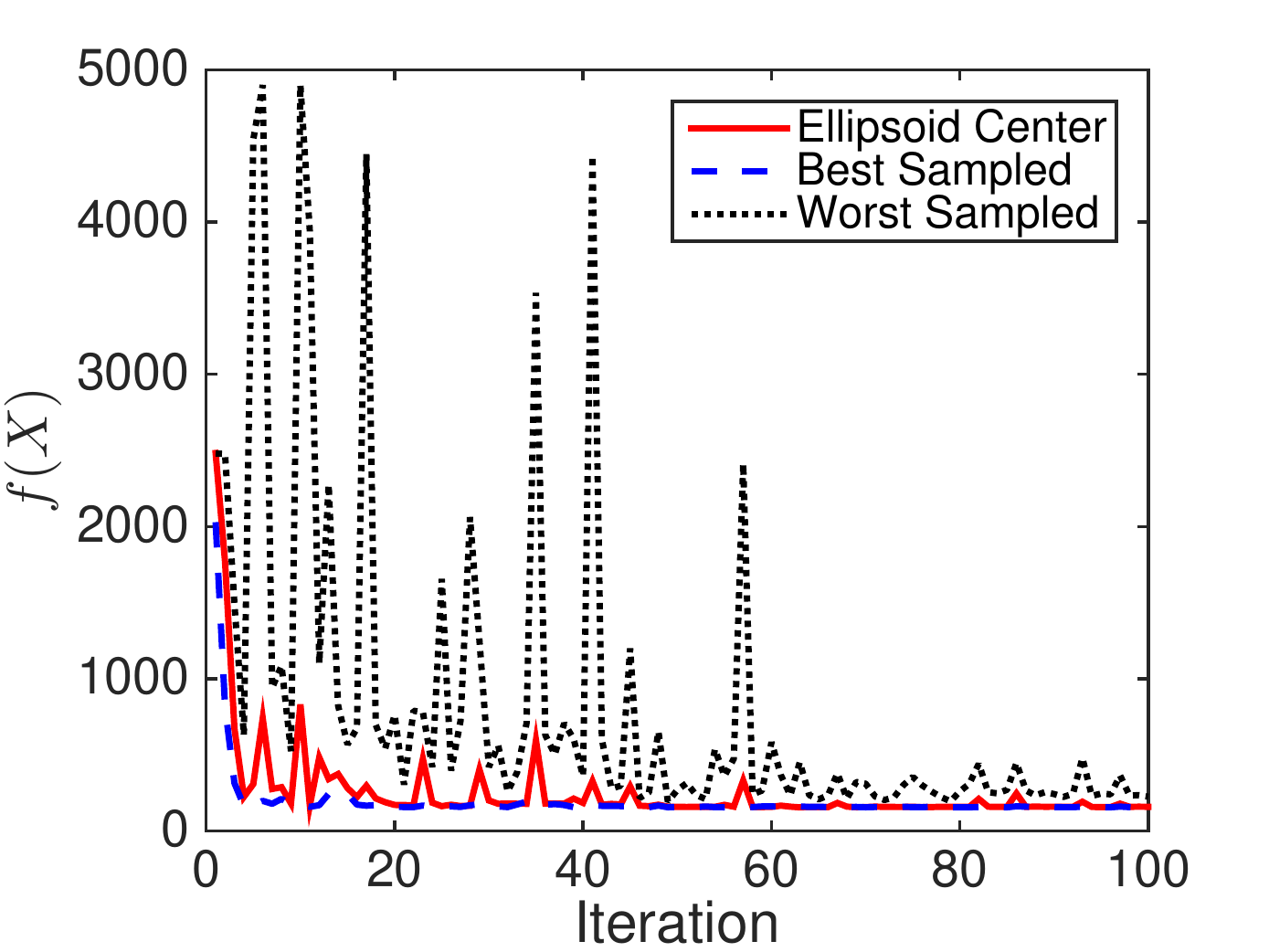}
\caption{Execution trace of QNSTOP}
\end{subfigure}
\hspace{2em}
\begin{subfigure}[]{0.45\textwidth}
\centering
\includegraphics[width=1\textwidth]{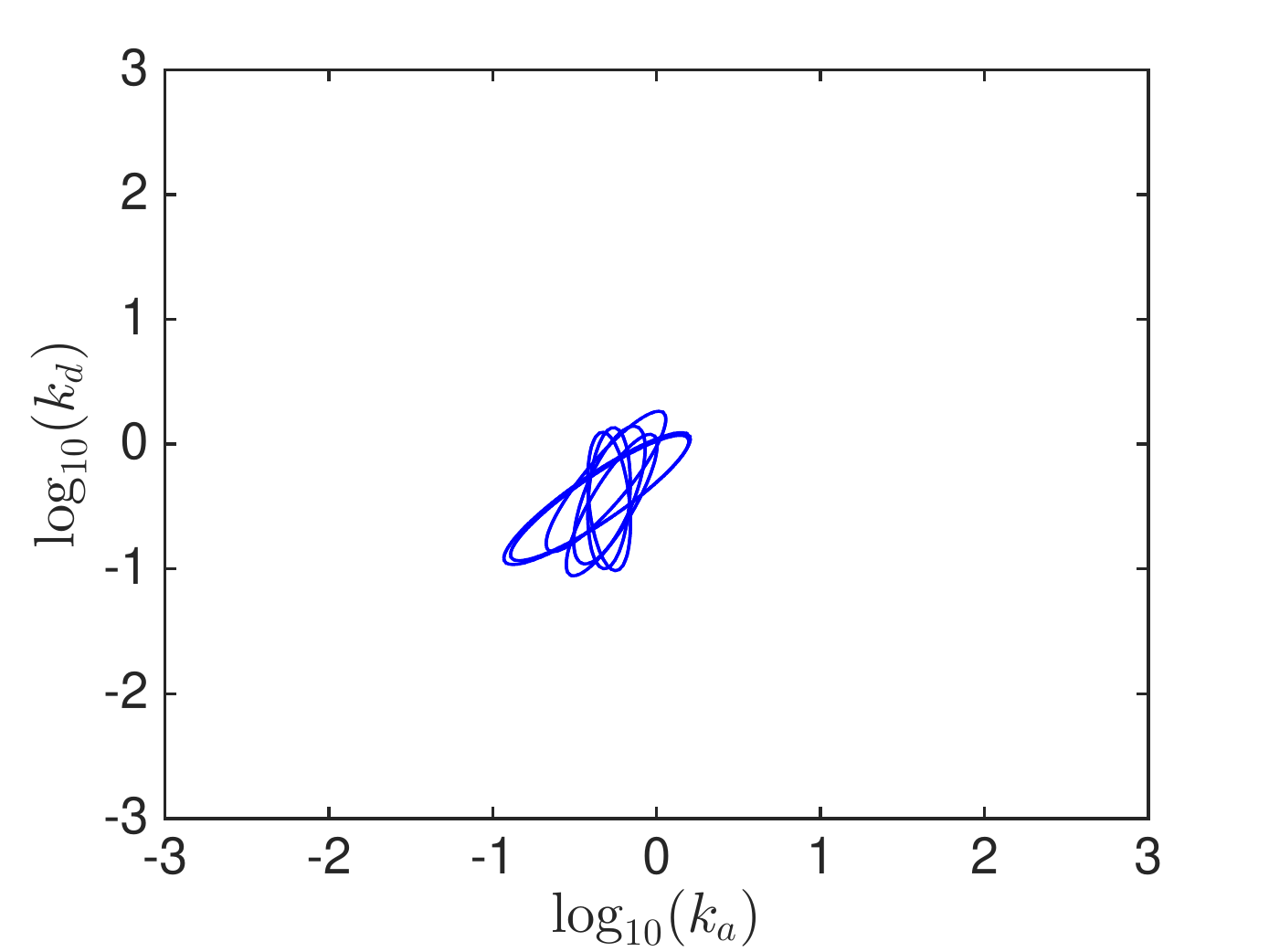}
\caption{Acceptable region}
\end{subfigure}
\begin{subfigure}[]{0.45\textwidth}
\centering
\includegraphics[width=1\textwidth]{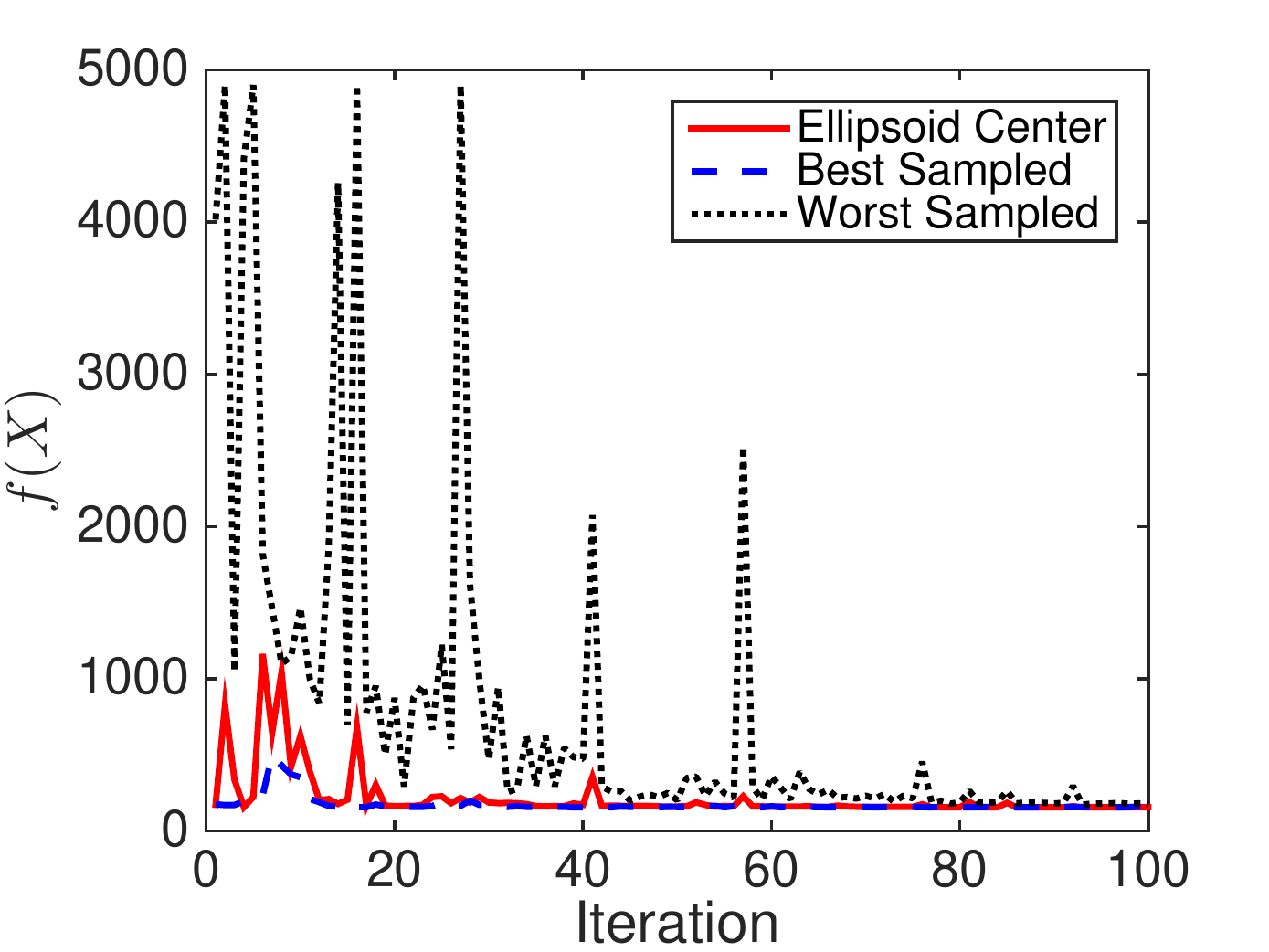}
\caption{Execution trace of QNSTOP}
\end{subfigure}
\hspace{2em}
\begin{subfigure}[]{0.45\textwidth}
\centering
\includegraphics[width=1\textwidth]{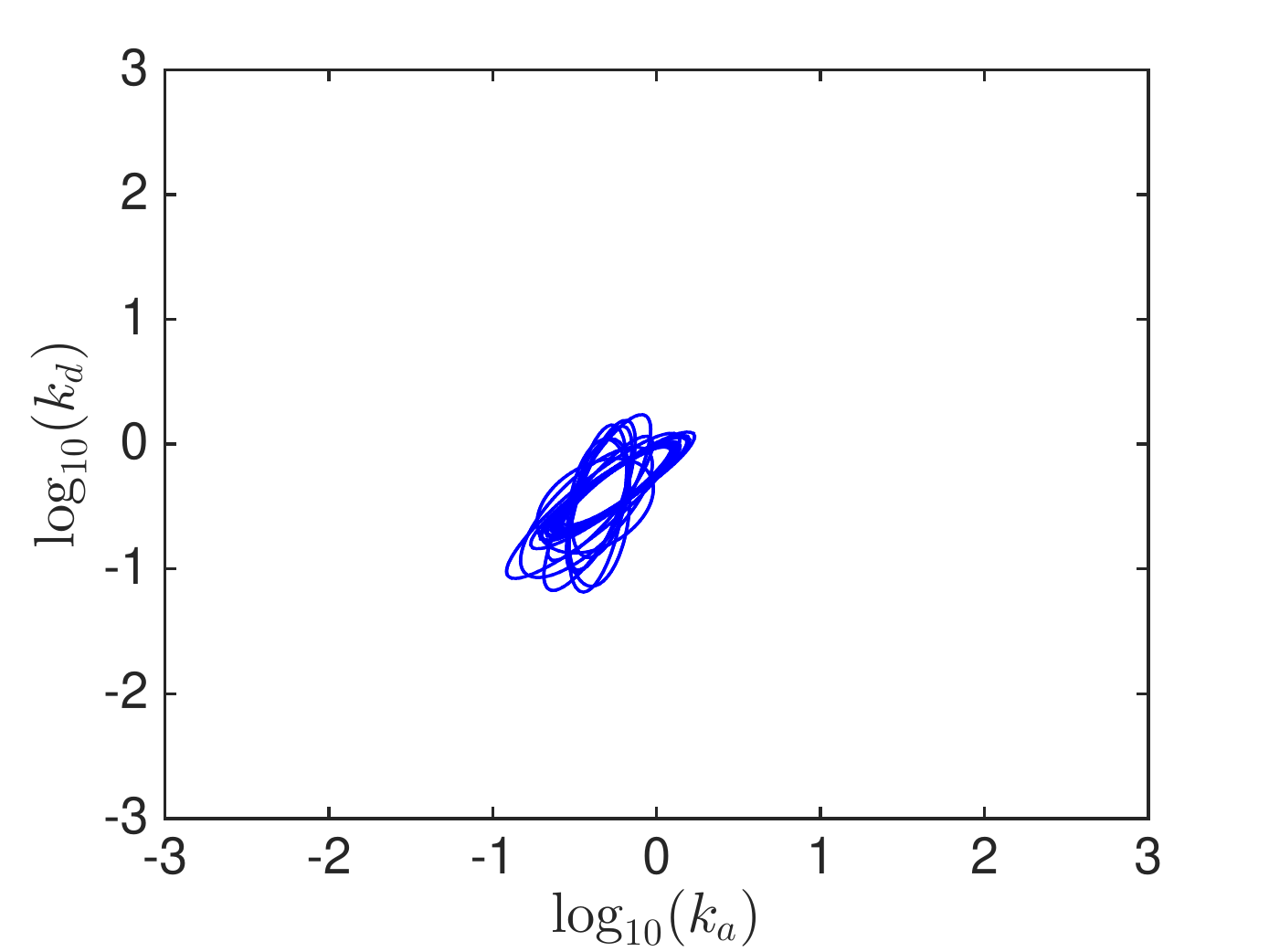}
\caption{Acceptable region}
\end{subfigure}
\begin{subfigure}[]{0.45\textwidth}
\centering
\includegraphics[width=1\textwidth]{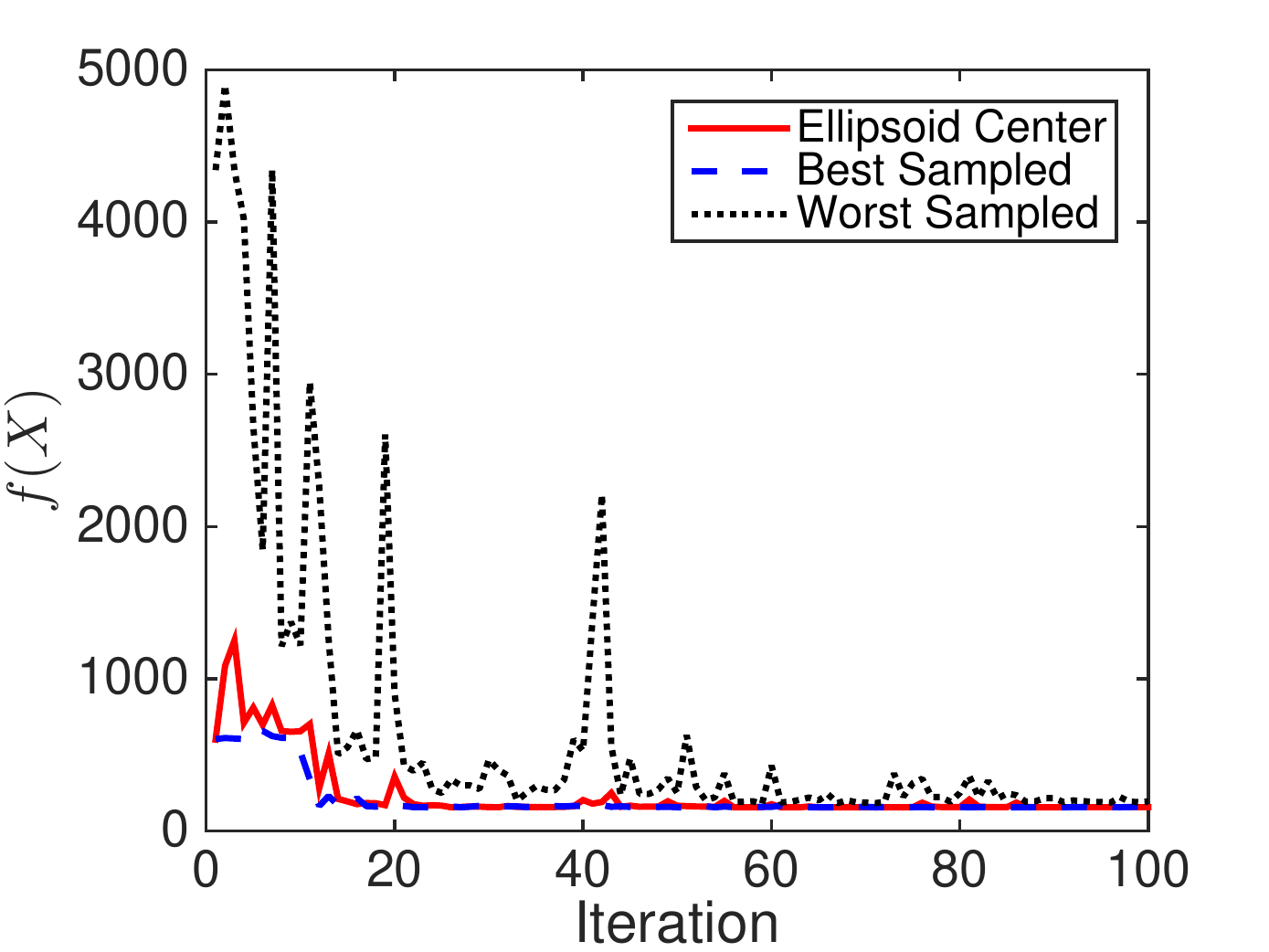}
\caption{Execution trace of QNSTOP}
\end{subfigure}
\hspace{2em}
\begin{subfigure}[]{0.45\textwidth}
\centering
\includegraphics[width=1\textwidth]{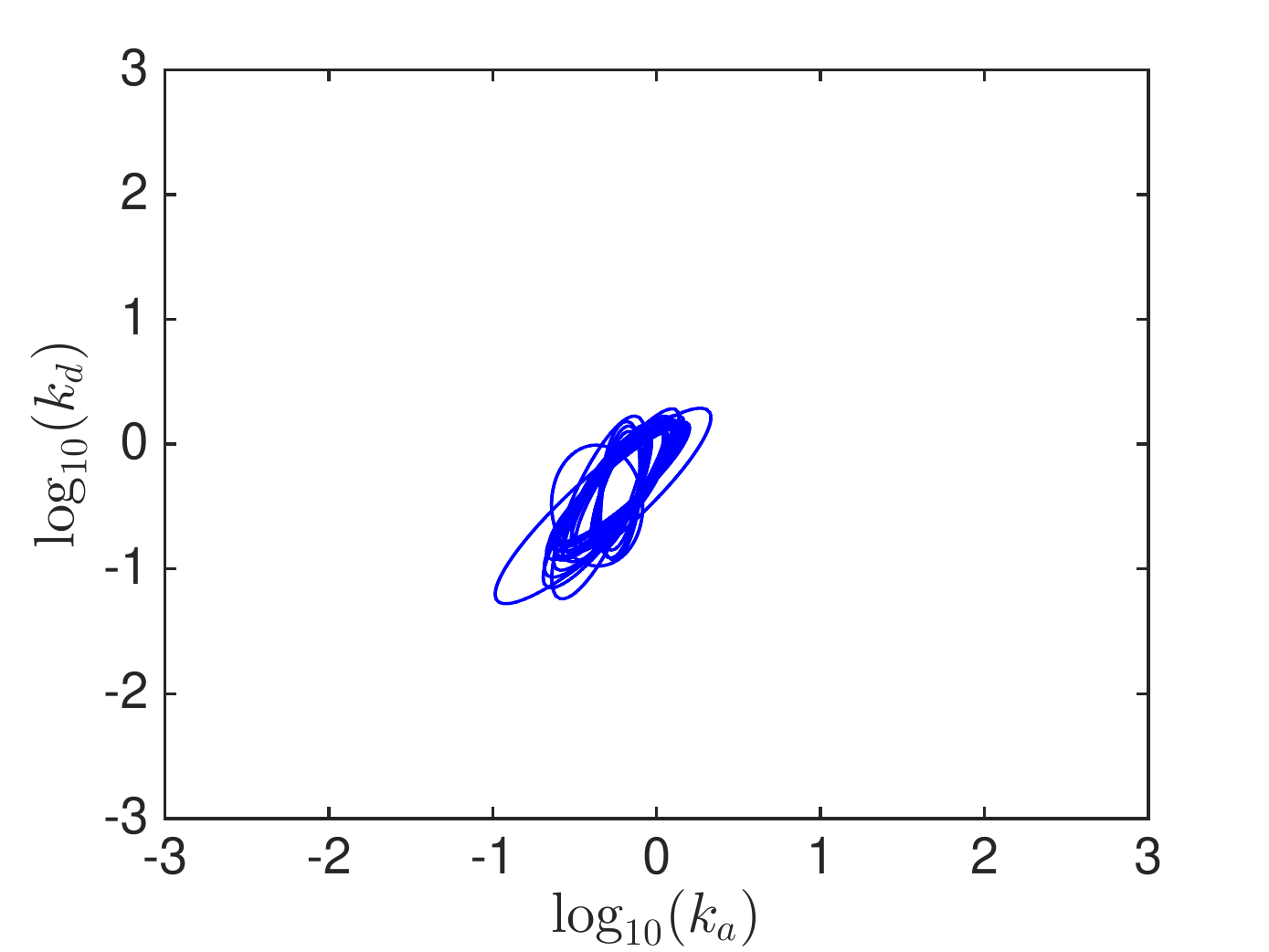}
\caption{Acceptable region}
\end{subfigure}
\caption{{Execution traces of QNSTOP (left column)
and acceptable regions of parameter space (right column) from maximum
log-likelihood method with different starting points}: (a, b) the lower
box corner $(\log_{10}(k_a), \log_{10}(k_d))=(-3,-3)$; (c, d) box center
$(\log_{10}(k_a), \log_{10}(k_d))=(0,0)$; (e, f) the upper box corner
$(\log_{10}(k_a), \log_{10}(k_d))=(3,3)$. The three types of execution
traces are: objective function values of the ellipsoid center at each
iteration (red line); minimum objective function values sampled in the
ellipsoidal region of each iteration (dashed blue line); maximum objective
function values sampled in the ellipsoidal region of each iteration (dotted
black line).} 
\label{fig:diff_points} 
\end{figure}

\textbf{Influence of starting points.} Fig.~\ref{fig:diff_points} shows
the execution traces of QNSTOP and the corresponding acceptable parameter
regions from different starting points: lower boundary point $L$, upper
boundary point $U$, and center point $(L+U)/2$. There are three types of
execution traces in Fig.~\ref{fig:diff_points}(a, c, e): the solid red
line is the objective function values of the ellipsoid center at each
iteration; the dashed blue line is the minimum objective function values
sampled in the ellipsoidal region of each iteration; the dotted black line
is the maximum objective function values sampled in the ellipsoidal region
of each iteration. From the execution traces of all three starting points,
the best sampled objective function value $f_l(X)$ decreases fast in the
first 20 iterations and then oscillates around 200. The worst sampled
objective function value also oscillates around 200 after 60 iterations.
In Fig.~\ref{fig:diff_points}(b, d, f), all ellipsoids satisfying the
$\alpha$-$\beta$-$\gamma$ rule, which form the acceptable region of parameter
space, are plotted in the domain. The acceptable parameter regions are
similar for all three methods and are located in the range of $[-1, 0.5]$
for both $\log_{10}(k_a)$ and $\log_{10}(k_d)$, regardless of the subtle
differences in region size and number of acceptable ellipsoids. Since
QNSTOP can choose multiple random starting points from a Latin Hypercube
experimental design, the rest of the paper shows the acceptable parameter
regions collected from 20 such starting points.

\begin{figure}[!htb]
\centering
\begin{subfigure}[]{0.45\textwidth}
\centering
\includegraphics[width=1\textwidth]{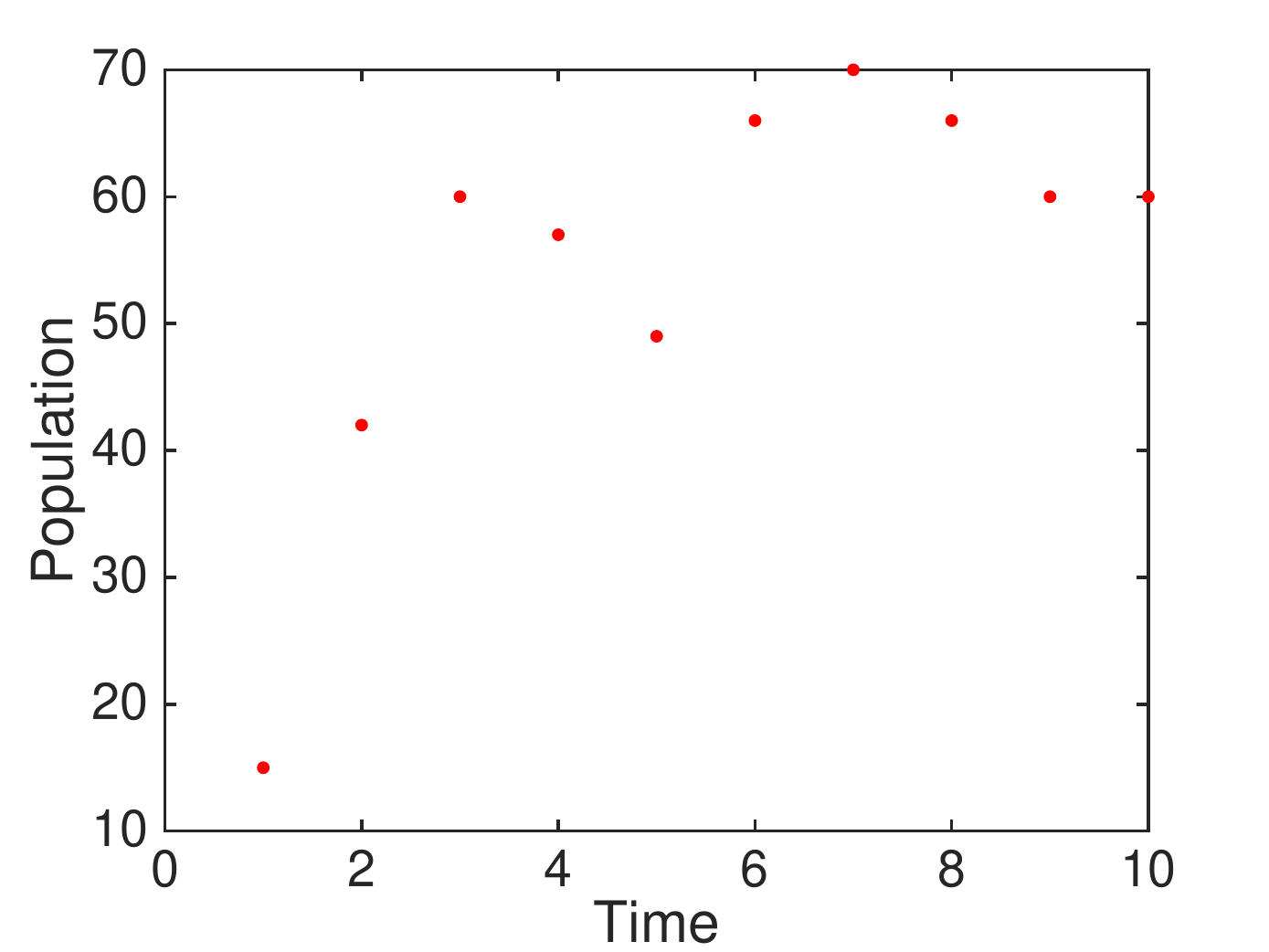}
\caption{Empirical data of ${\rm B}_n$ population}
\end{subfigure}
\hspace{2em}
\begin{subfigure}[]{0.45\textwidth}
\centering
\includegraphics[width=1\textwidth]{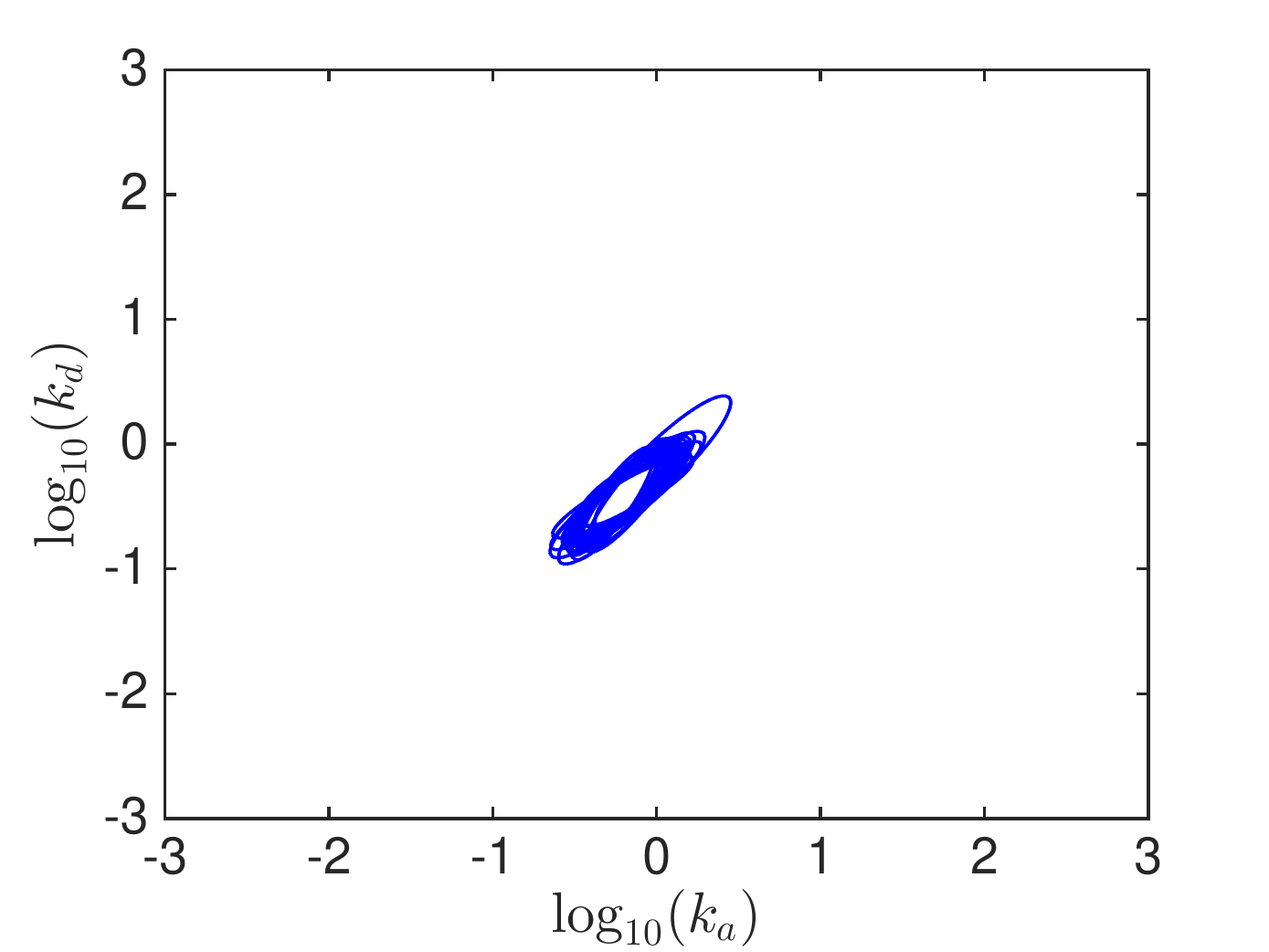}
\caption{Acceptable region}
\end{subfigure}
\begin{subfigure}[]{0.45\textwidth}
\centering
\includegraphics[width=1\textwidth]{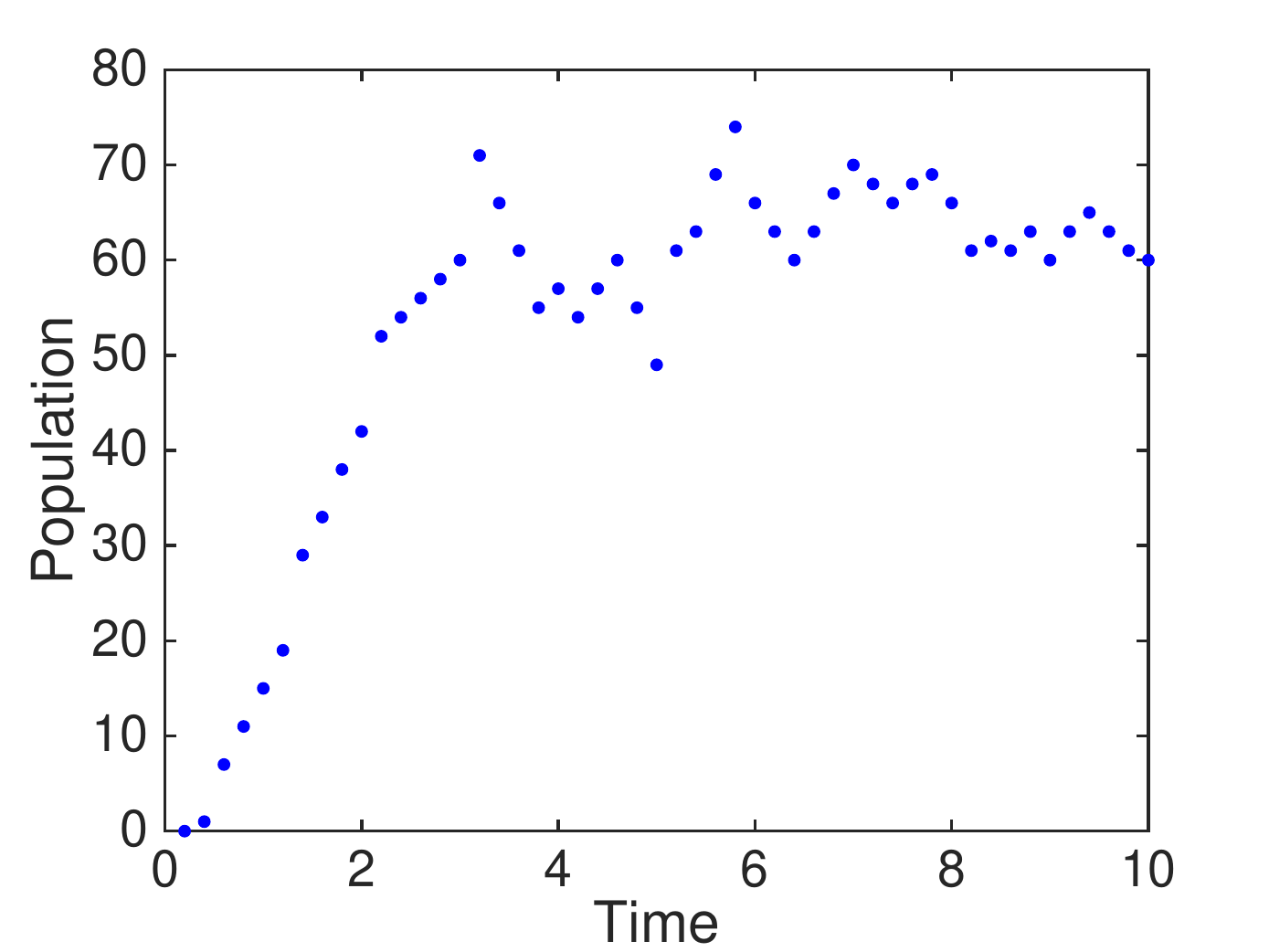}
\caption{Empirical data of ${\rm B}_n$ population}
\end{subfigure}
\hspace{2em}
\begin{subfigure}[]{0.45\textwidth}
\centering
\includegraphics[width=1\textwidth]{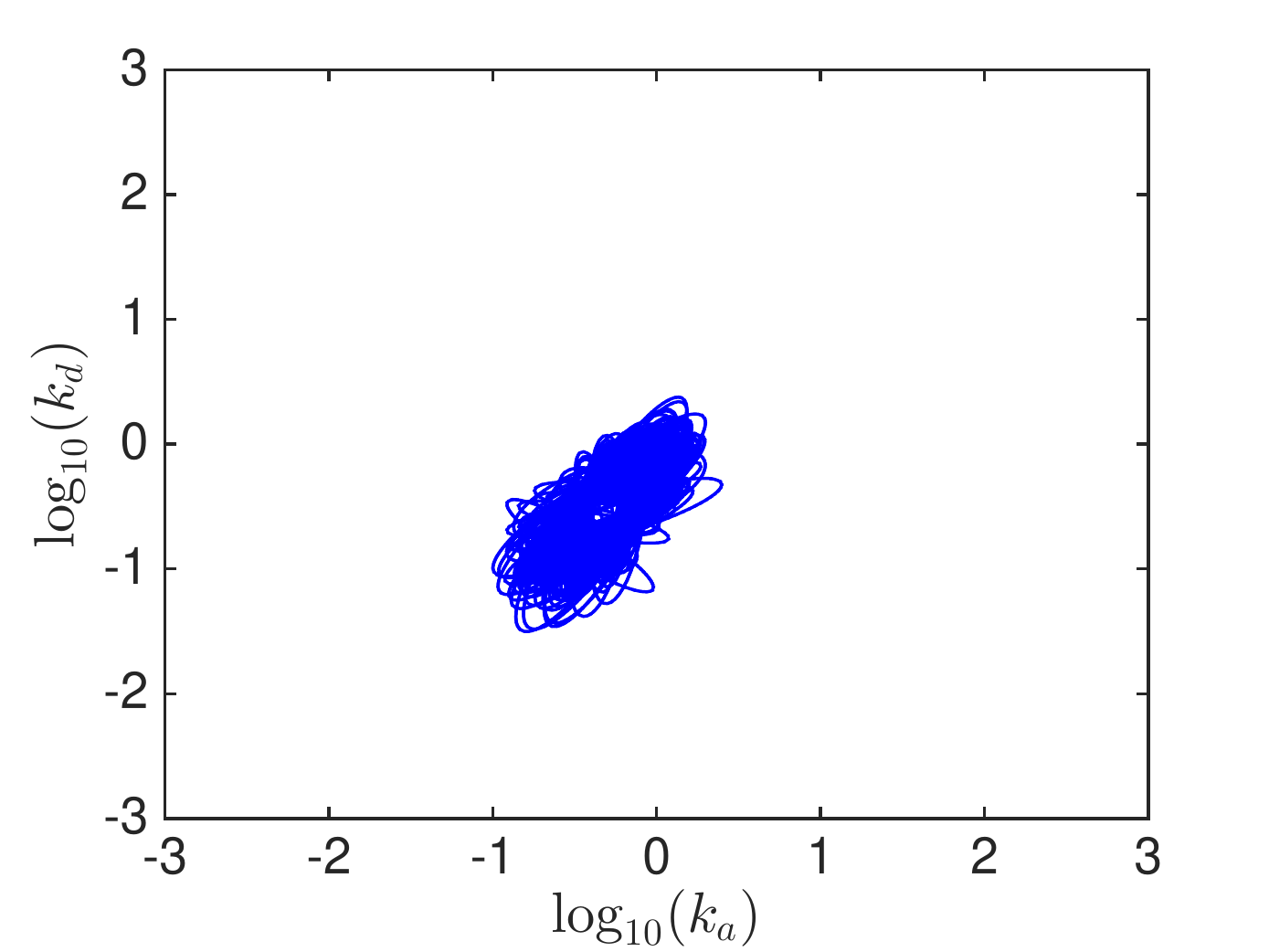}
\caption{Acceptable region}
\end{subfigure}
\begin{subfigure}[]{0.45\textwidth}
\centering
\includegraphics[width=1\textwidth]{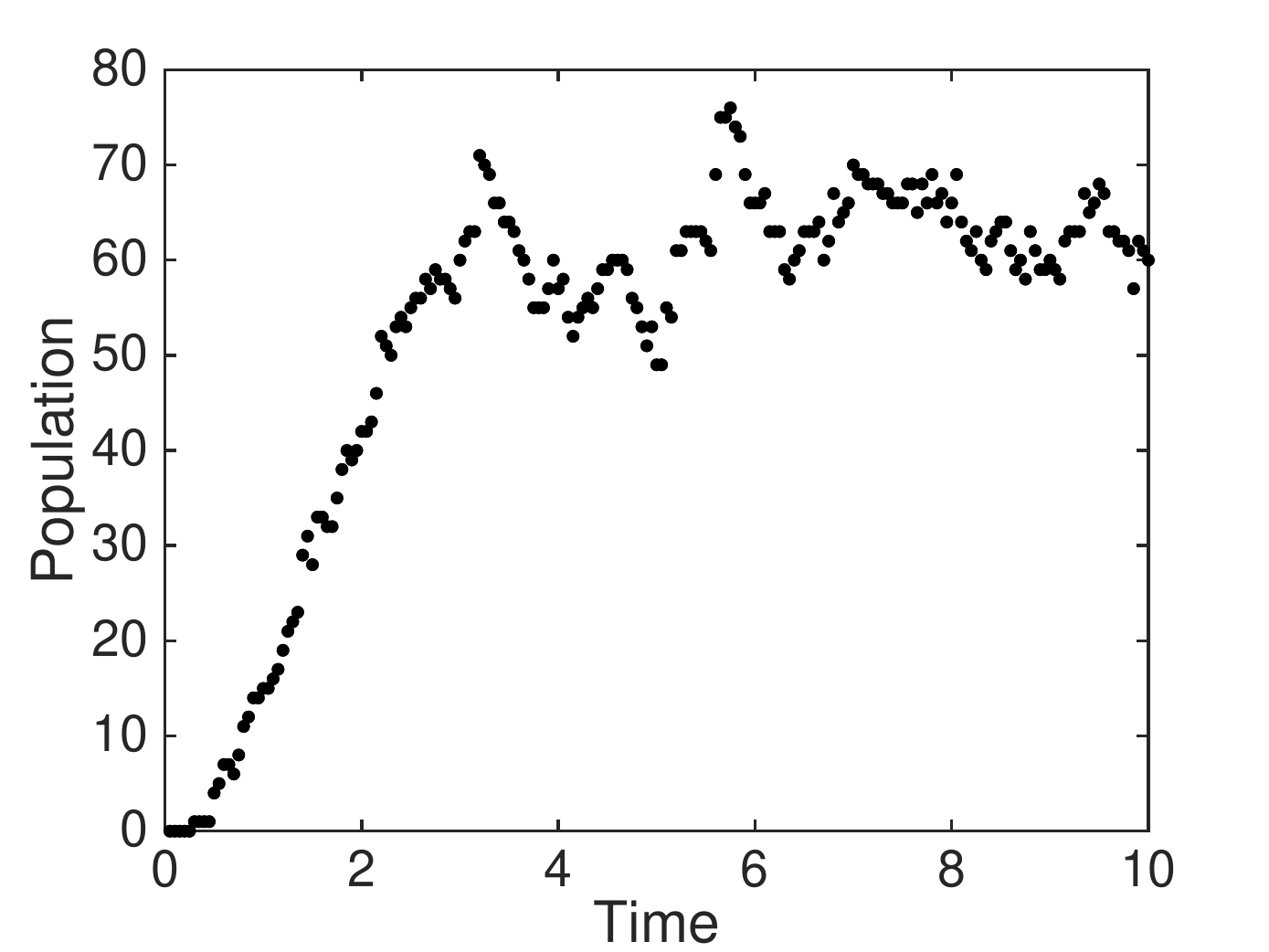}
\caption{Empirical data of ${\rm B}_n$ population}
\end{subfigure}
\hspace{2em}
\begin{subfigure}[]{0.45\textwidth}
\centering
\includegraphics[width=1\textwidth]{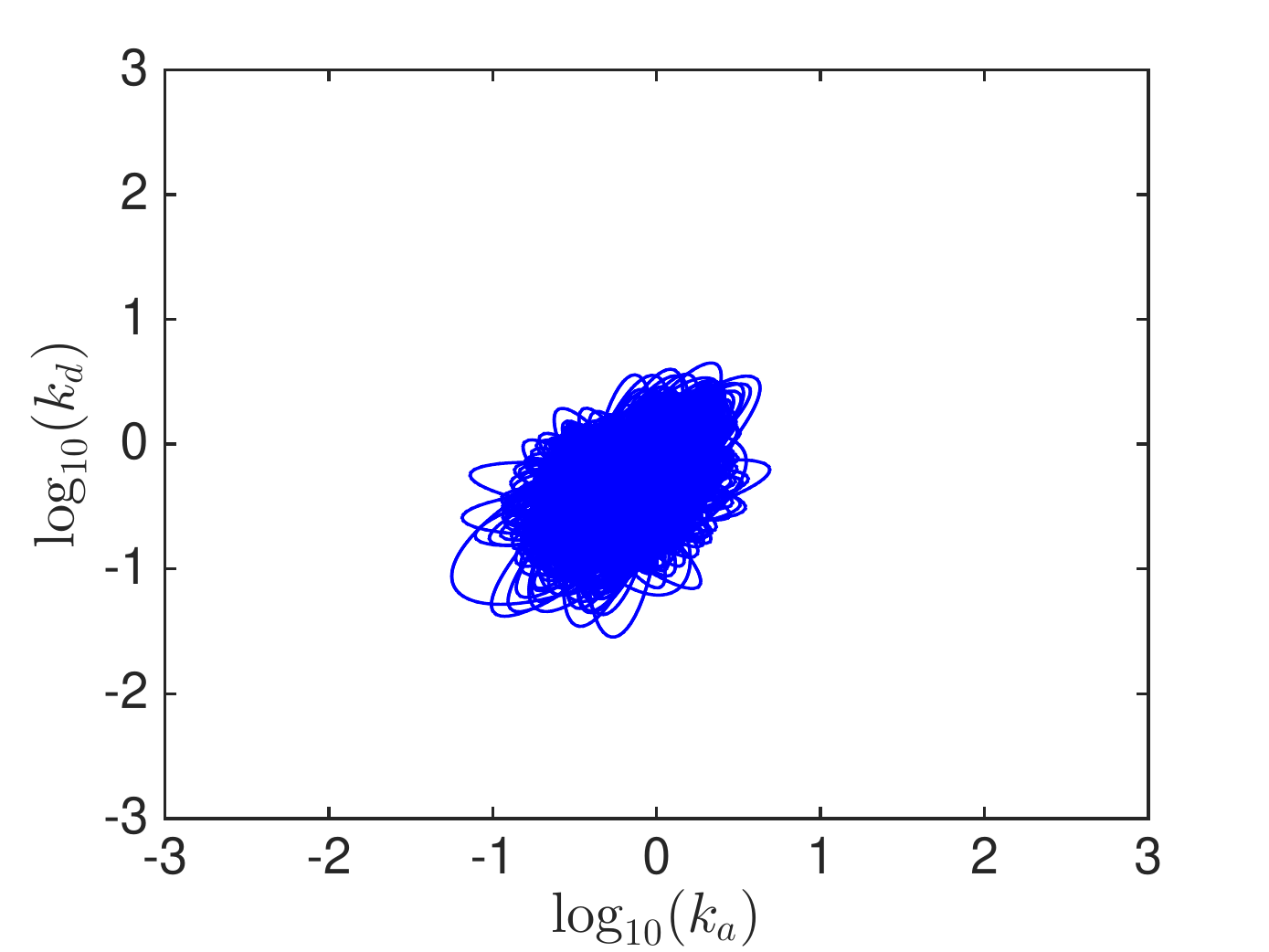}
\caption{Acceptable region}
\end{subfigure}
\caption{{Different empirical data of ${\rm B}_n$ population (left column)
and the corresponding acceptable parameter region (right column) from
maximum log-likelihood method. The empirical data are sampled using
different time steps $\tau$ and sample sizes $m$}: (a, b) $\tau=1$, $m=10$;
(c, d) $\tau=0.2$, $m=50$; (e, f) $\tau=0.05$, $m=200$. The acceptable
parameter region is the union of results from 20 starting points.}
\label{fig:diff_tau}
\end{figure}

\textbf{Influence of empirical data.} To study the robustness of the
algorithm, vary the empirical data and the data size. Fig.~\ref{fig:diff_tau}
illustrates the acceptable regions from the same population trajectory of
${\rm B}_n$ but with different numbers of data points.
Fig.~\ref{fig:diff_tau}(a) samples ten data points of ${\rm B}_n$ population
with a time step $\tau=1$, while Fig.~\ref{fig:diff_tau}(b),(c) sample 50
and 200 data points with time steps $\tau=0.2$ and $\tau=0.05$, respectively.
The acceptable region of parameter space increases with the data size $m$.

\begin{figure}[!htb]
\centering
\begin{subfigure}[]{0.45\textwidth}
\centering
\includegraphics[width=1\textwidth]{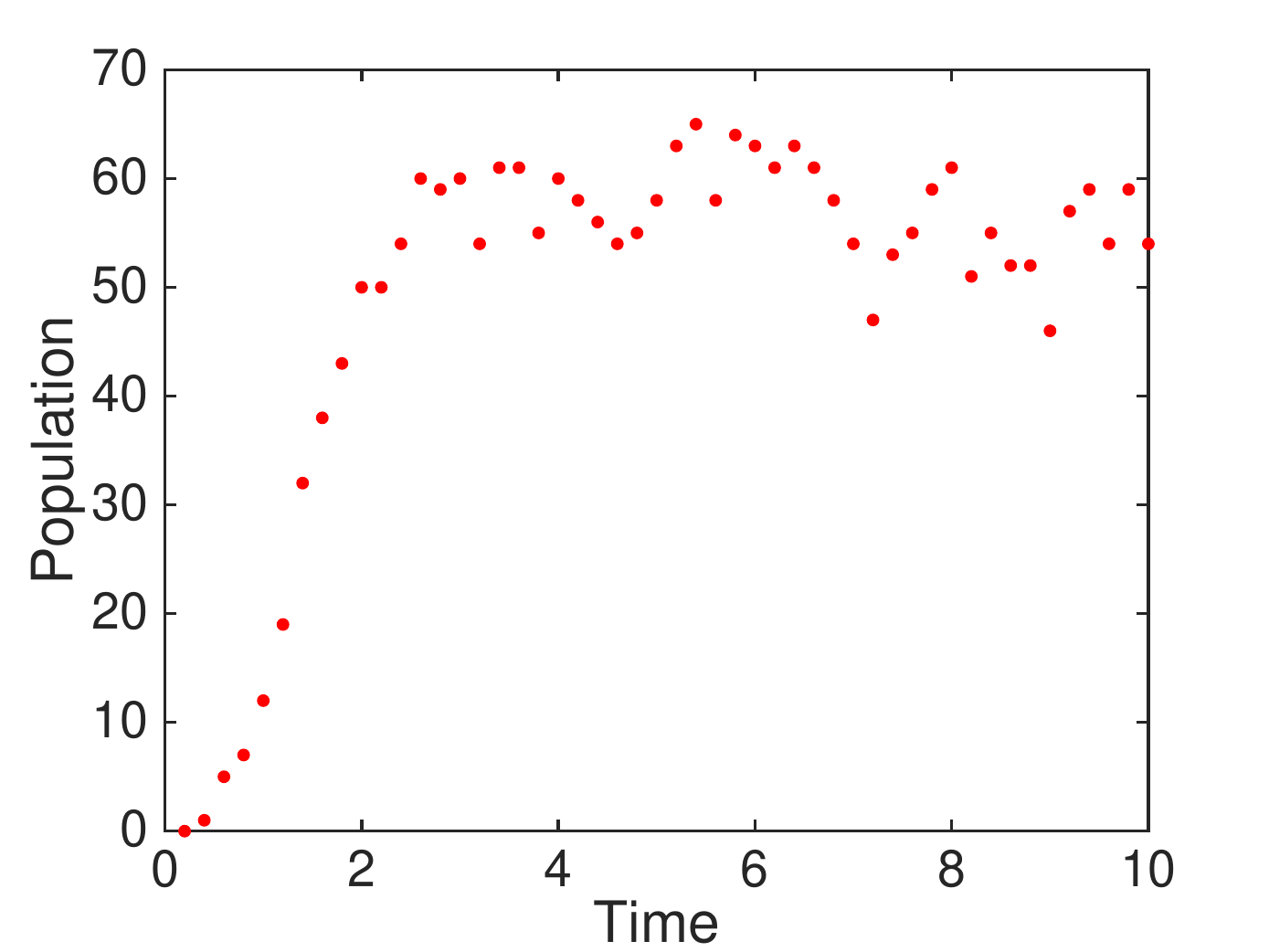}
\caption{Empirical data of ${\rm B}_n$ population}
\end{subfigure}
\hspace{2em}
\begin{subfigure}[]{0.45\textwidth}
\centering
\includegraphics[width=1\textwidth]{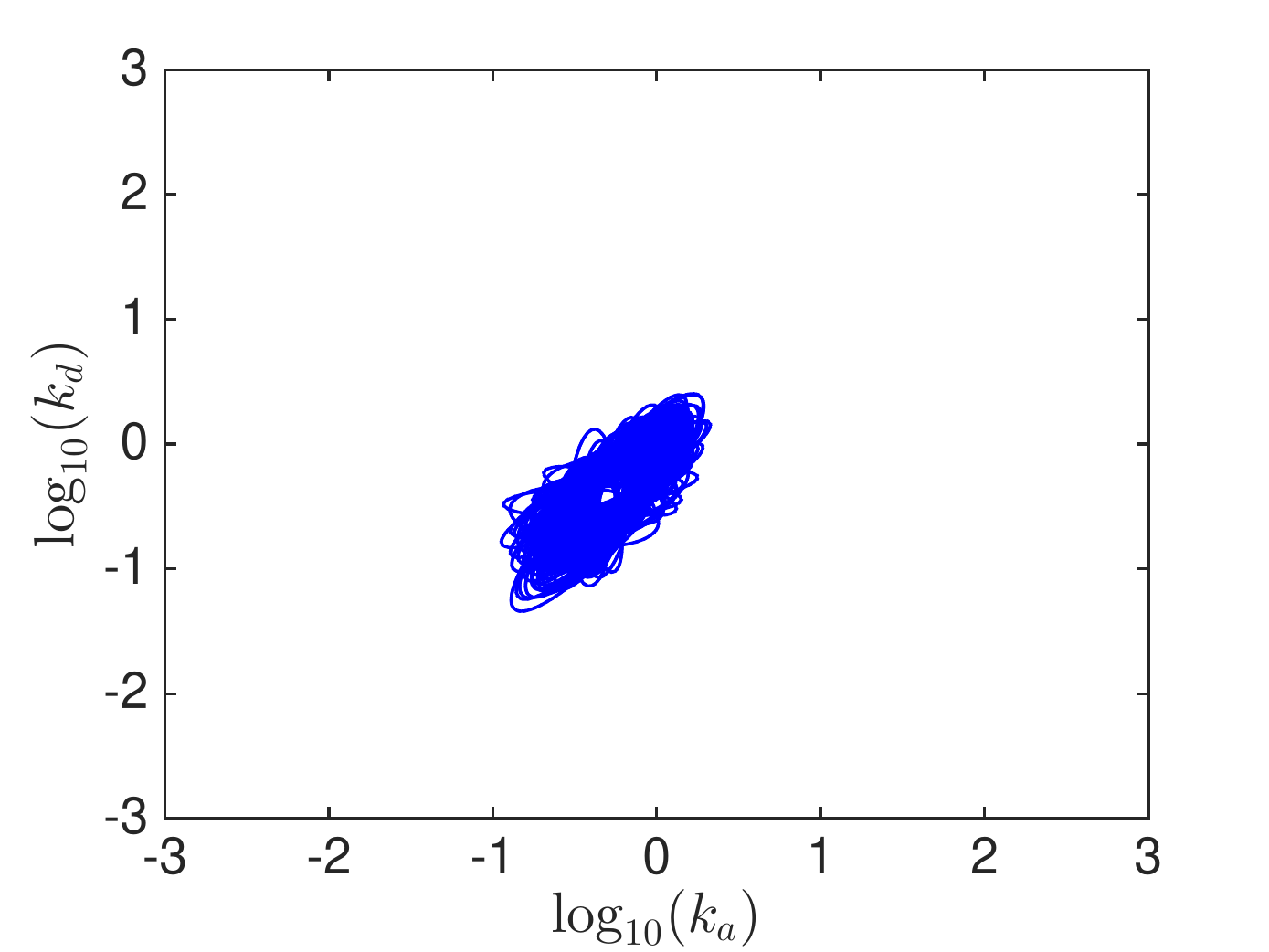}
\caption{Acceptable region}
\end{subfigure}
\begin{subfigure}[]{0.45\textwidth}
\centering
\includegraphics[width=1\textwidth]{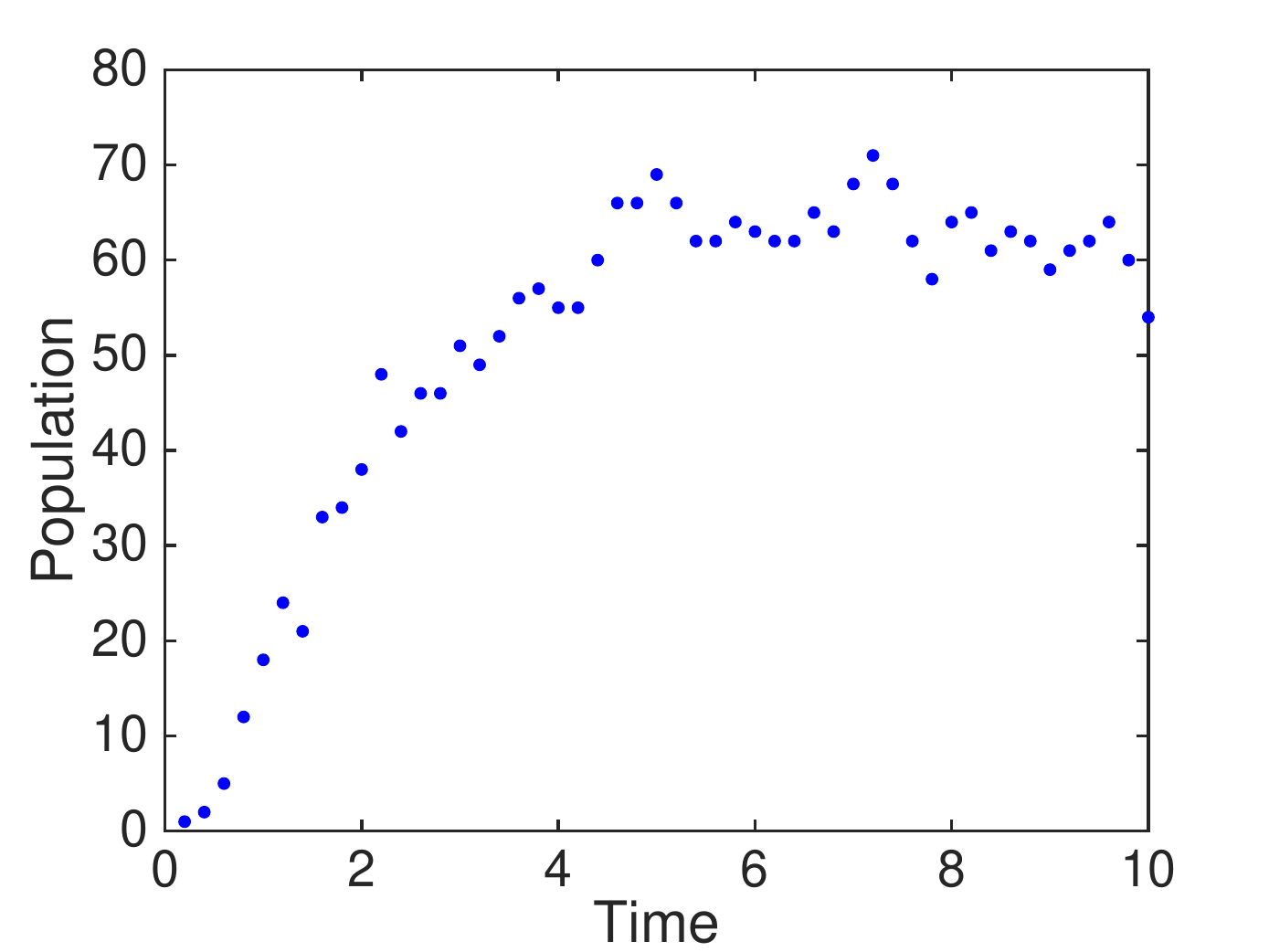}
\caption{Empirical data of ${\rm B}_n$ population}
\end{subfigure}
\hspace{2em}
\begin{subfigure}[]{0.45\textwidth}
\centering
\includegraphics[width=1\textwidth]{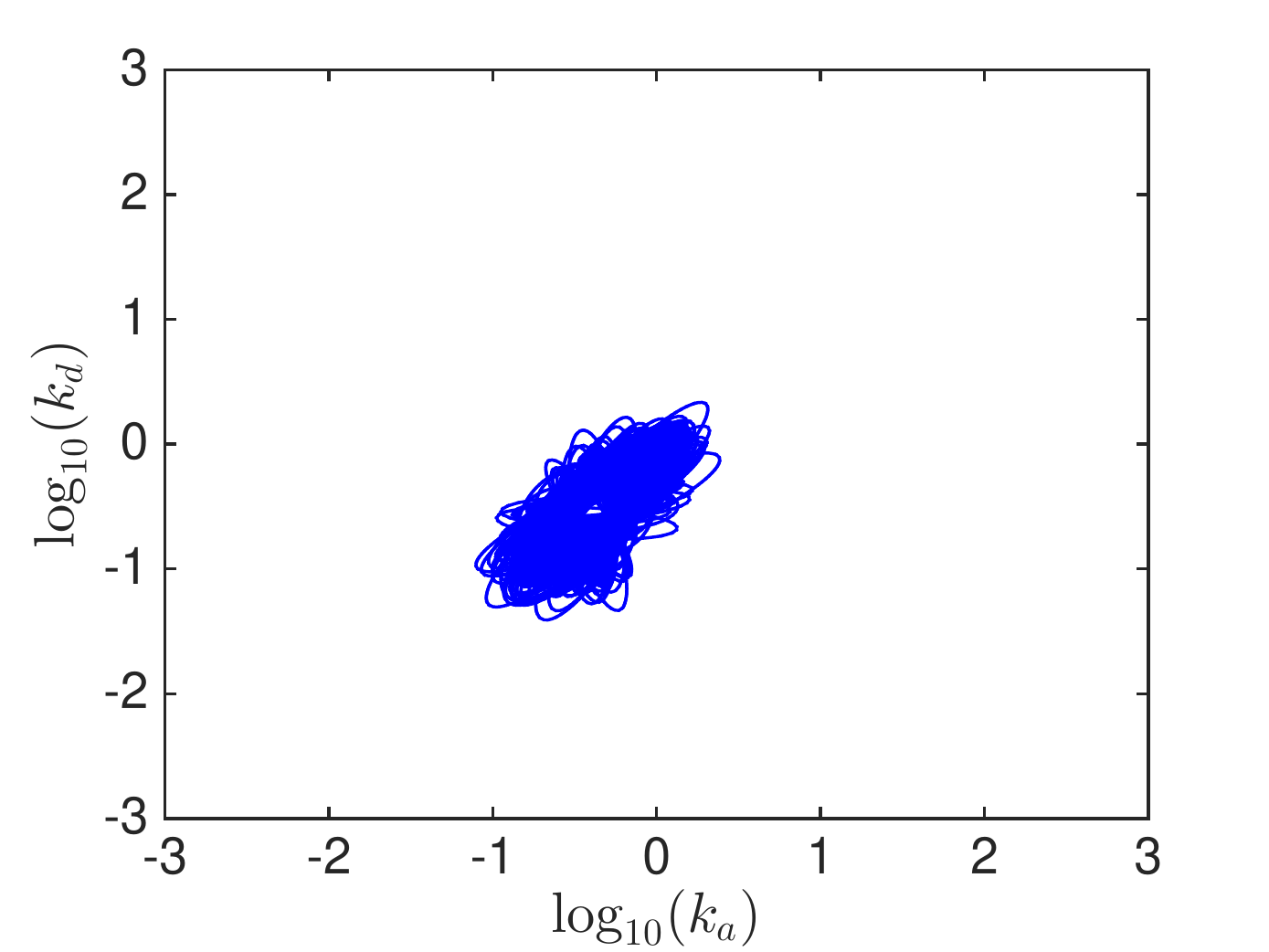}
\caption{Acceptable region}
\end{subfigure}
\begin{subfigure}[]{0.45\textwidth}
\centering
\includegraphics[width=1\textwidth]{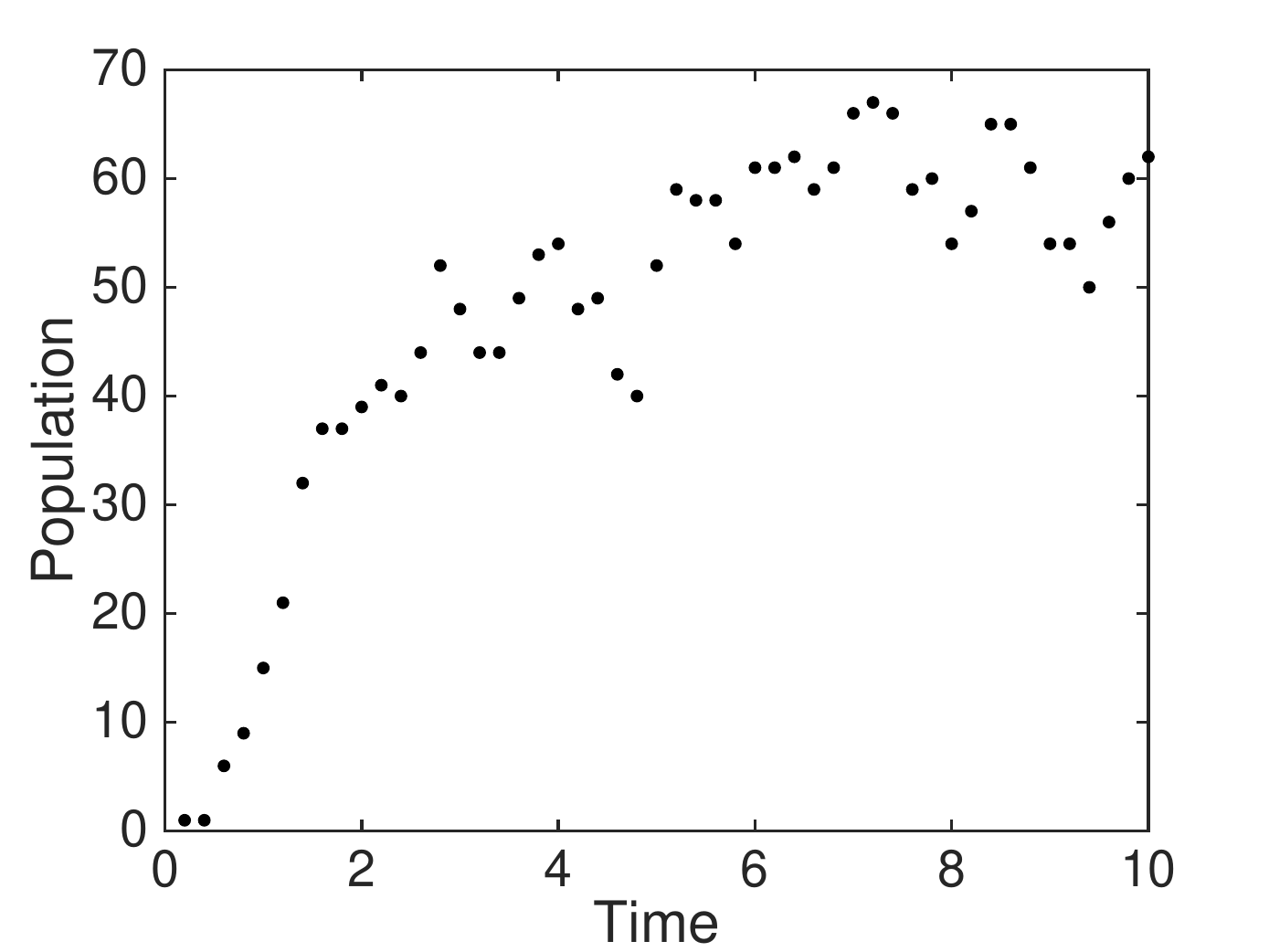}
\caption{Empirical data of ${\rm B}_n$ population}
\end{subfigure}
\hspace{2em}
\begin{subfigure}[]{0.45\textwidth}
\centering
\includegraphics[width=1\textwidth]{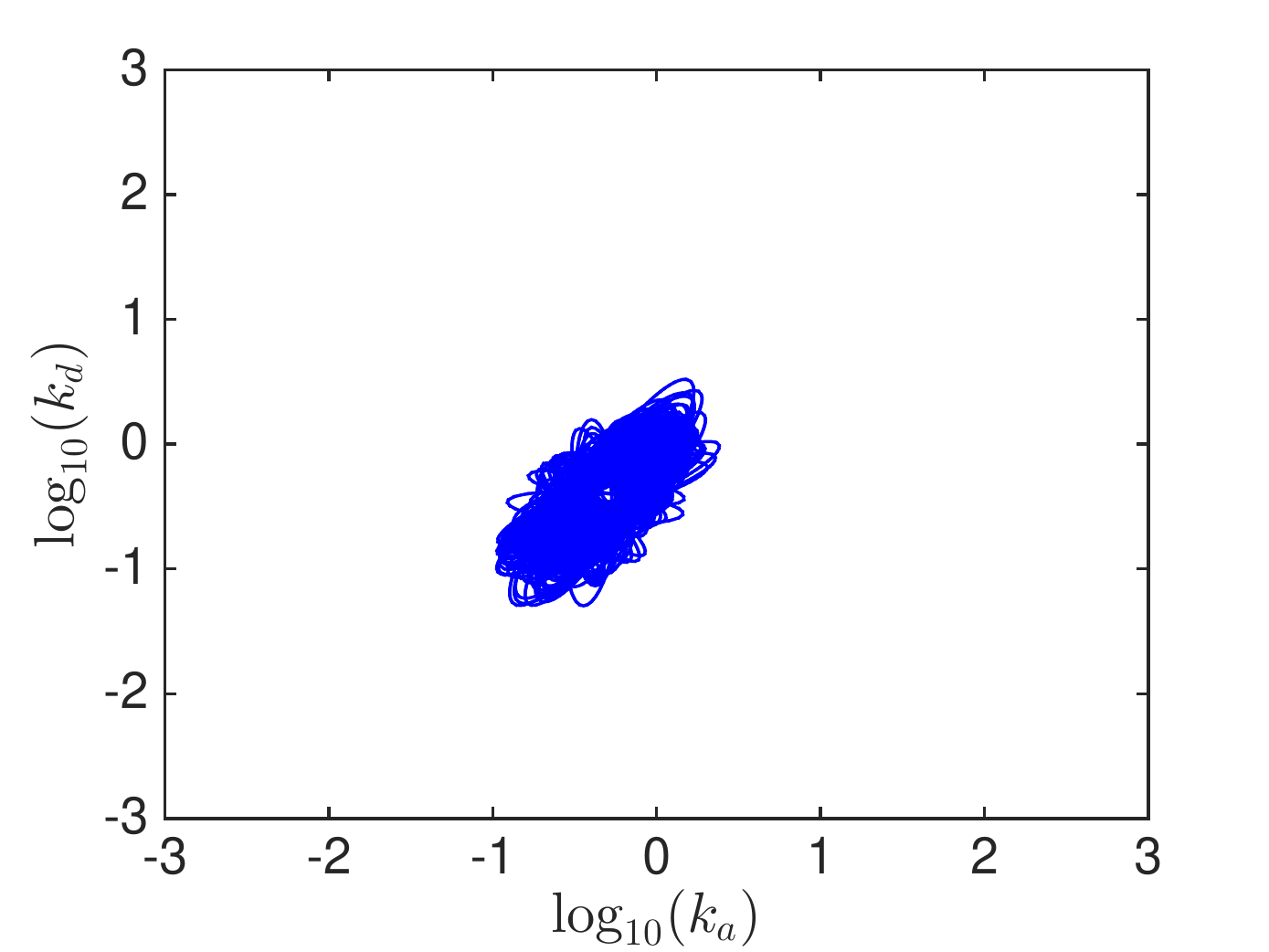}
\caption{Acceptable region}
\end{subfigure}
\caption{Optimization results of maximum log-likelihood with three different
empirical data sets containing 50 data points, collected every 0.2 time
unit. The acceptable region is the union of results from 20 starting points.}
\label{fig:diff_data}
\end{figure}

Fig.~\ref{fig:diff_data} shows the optimization results from three different
empirical datasets of ${\rm B}_n$ population trajectories sampled from the
stochastic multisite phosphorylation system. With different population
trajectories, the acceptable parameter regions are consistent in size and
shape. Thus, the parameter optimization of the stochastic Hill function
system is more sensitive to the size of dataset than the data content
variation.
					
\subsection{Full parameter case}
This subsection studies the full parameter vector, in which all four
parameters $k_a$, $k_d$, $\sigma$, and $k_m$ are unknown in the stochastic
Hill function system. The QNSTOP settings are the same except the initial
ellipsoid radius $\hbox{TAU}=1.3$ and the sample point number $N=20$. The
empirical data is one simulated trajectory of the ${\rm B}_n$ population
where $m=50$, $\tau=0.2$. The values of $\alpha\hbox{-}\beta\hbox{-}\gamma$
defining the acceptable regions are set differently according to the
objective functions.

\begin{figure}[!htb]
\centering
\begin{subfigure}[]{0.45\textwidth}
\centering
\includegraphics[width=1\textwidth]{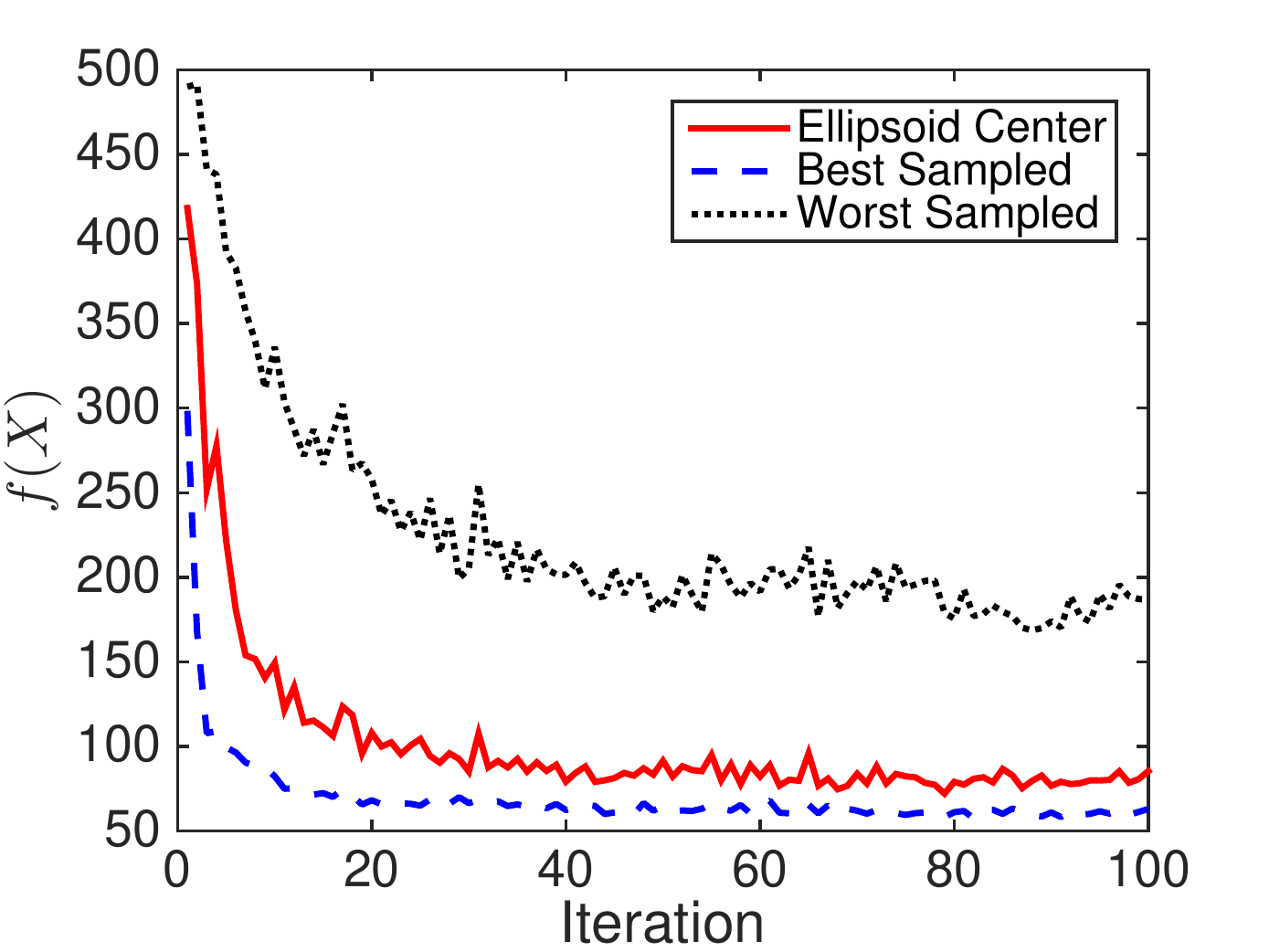}
\caption{Average execution trace of QNSTOP}
\end{subfigure}
\hspace{2em}
\begin{subfigure}[]{0.45\textwidth}
\centering
\includegraphics[width=1\textwidth]{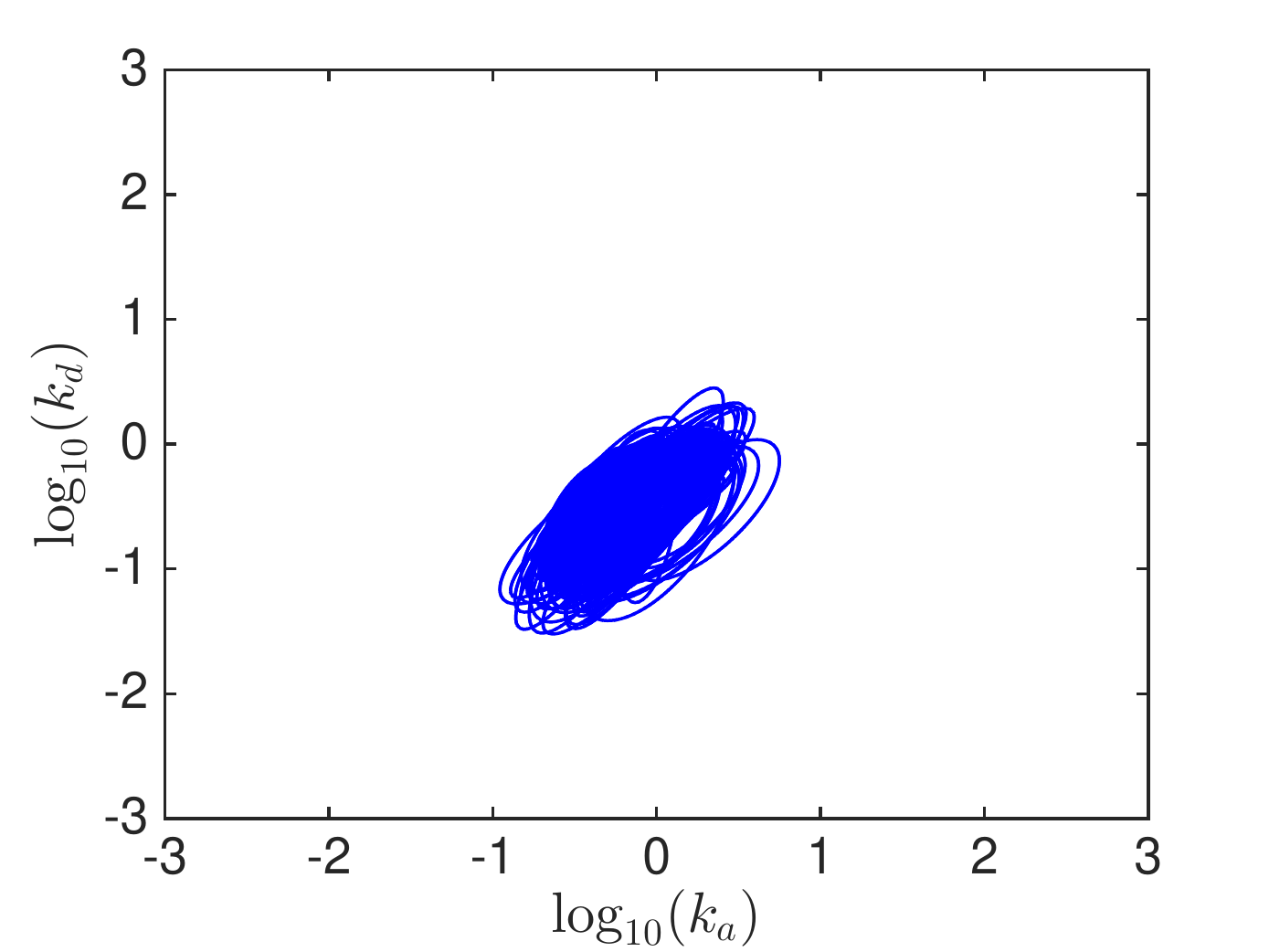}
\caption{Acceptable region}
\end{subfigure}
\begin{subfigure}[]{0.45\textwidth}
\centering
\includegraphics[width=1\textwidth]{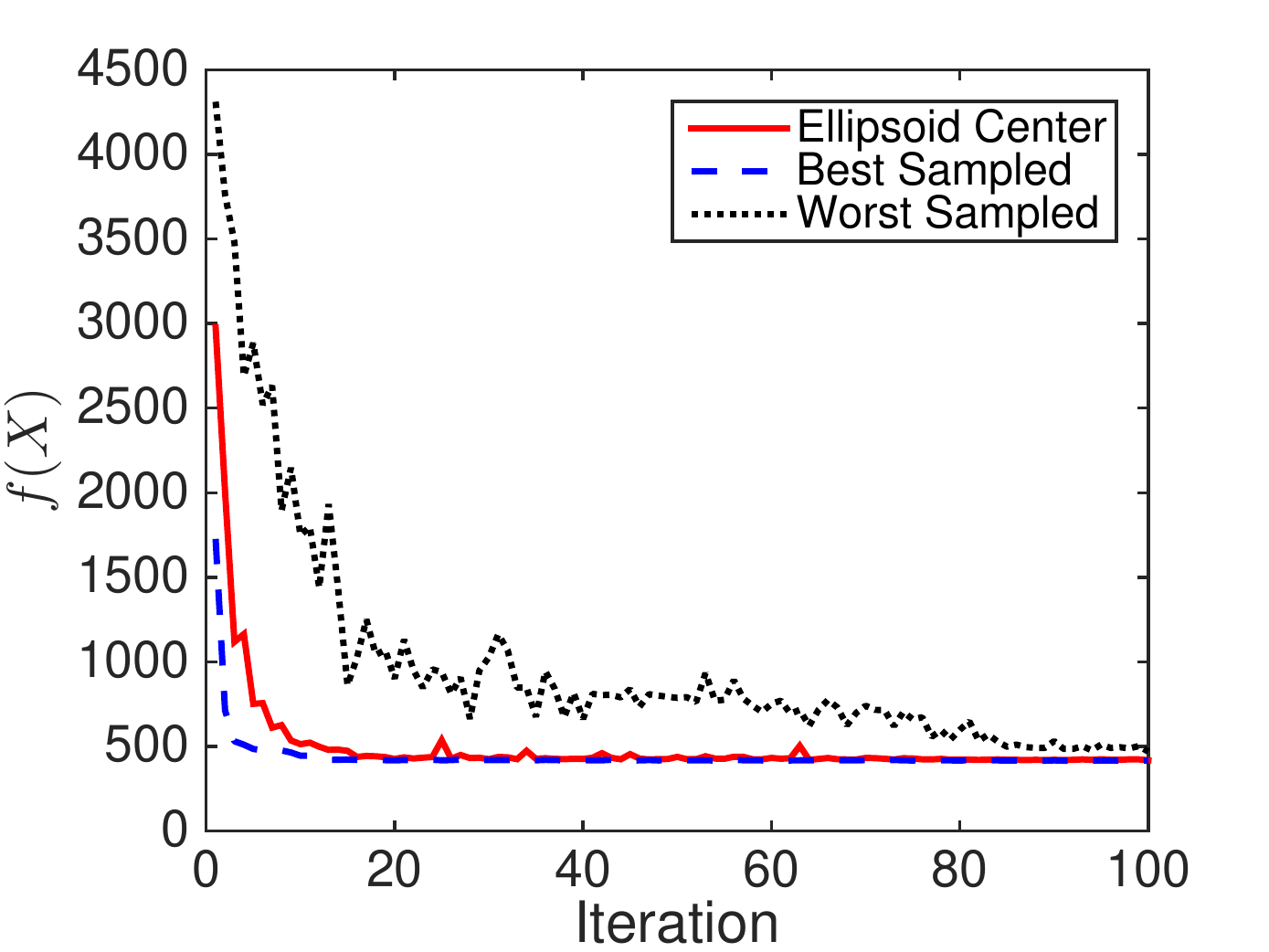}
\caption{Average execution trace of QNSTOP}
\end{subfigure}
\hspace{2em}
\begin{subfigure}[]{0.45\textwidth}
\centering
\includegraphics[width=1\textwidth]{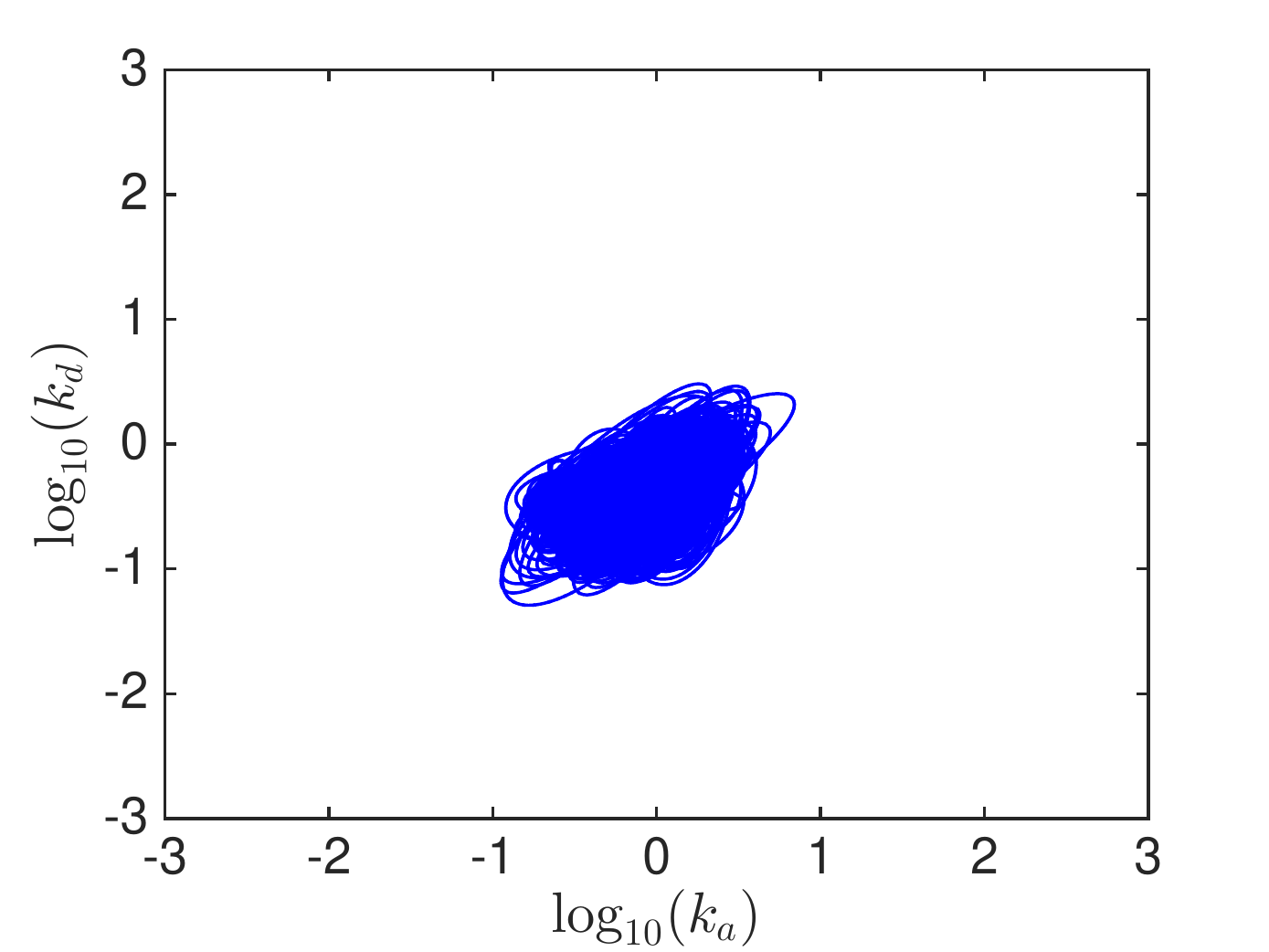}
\caption{Acceptable region}
\end{subfigure}
\begin{subfigure}[]{0.45\textwidth}
\centering
\includegraphics[width=1\textwidth]{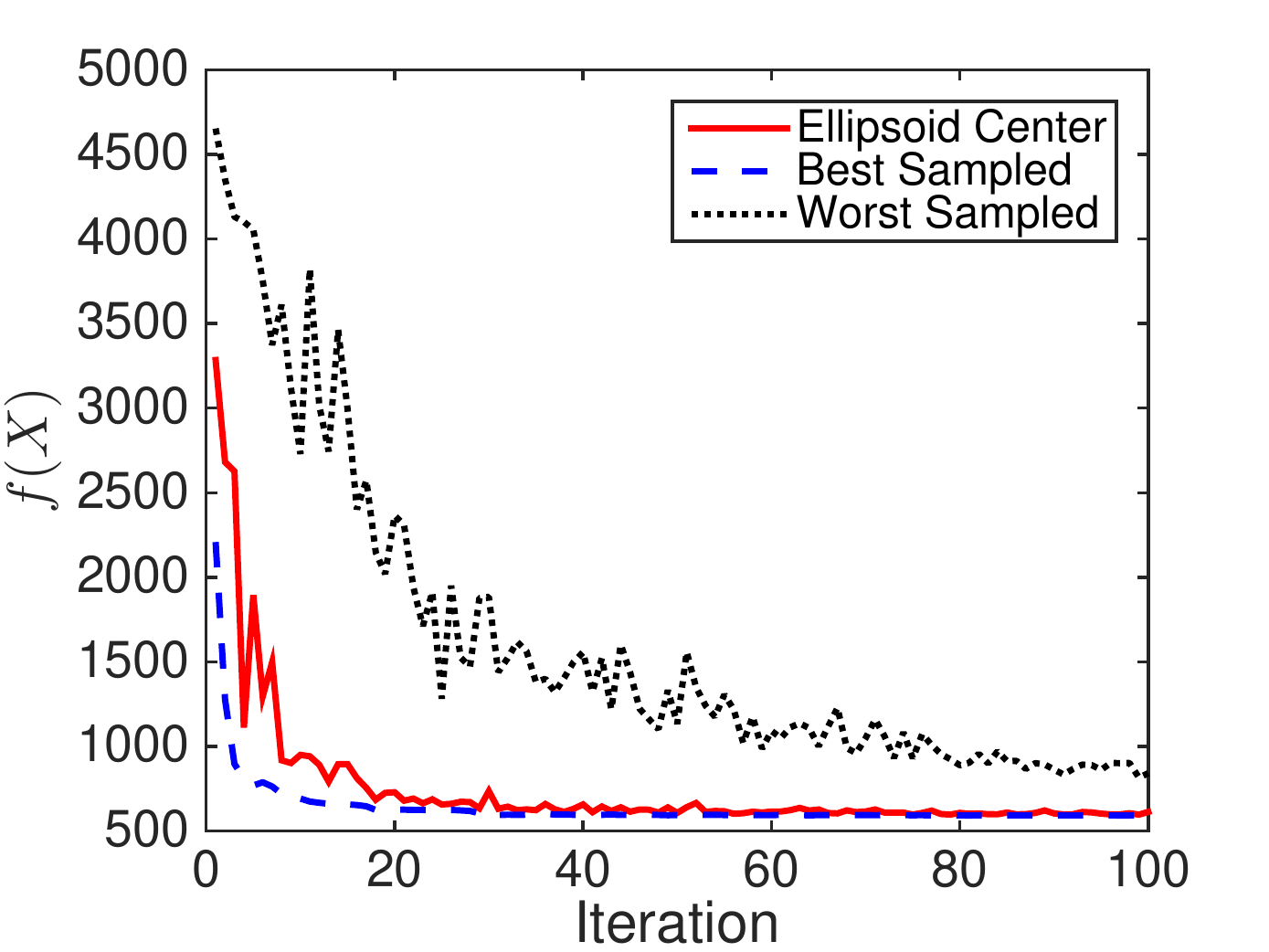}
\caption{Average execution trace of QNSTOP}
\end{subfigure}
\hspace{2em}
\begin{subfigure}[]{0.45\textwidth}
\centering
\includegraphics[width=1\textwidth]{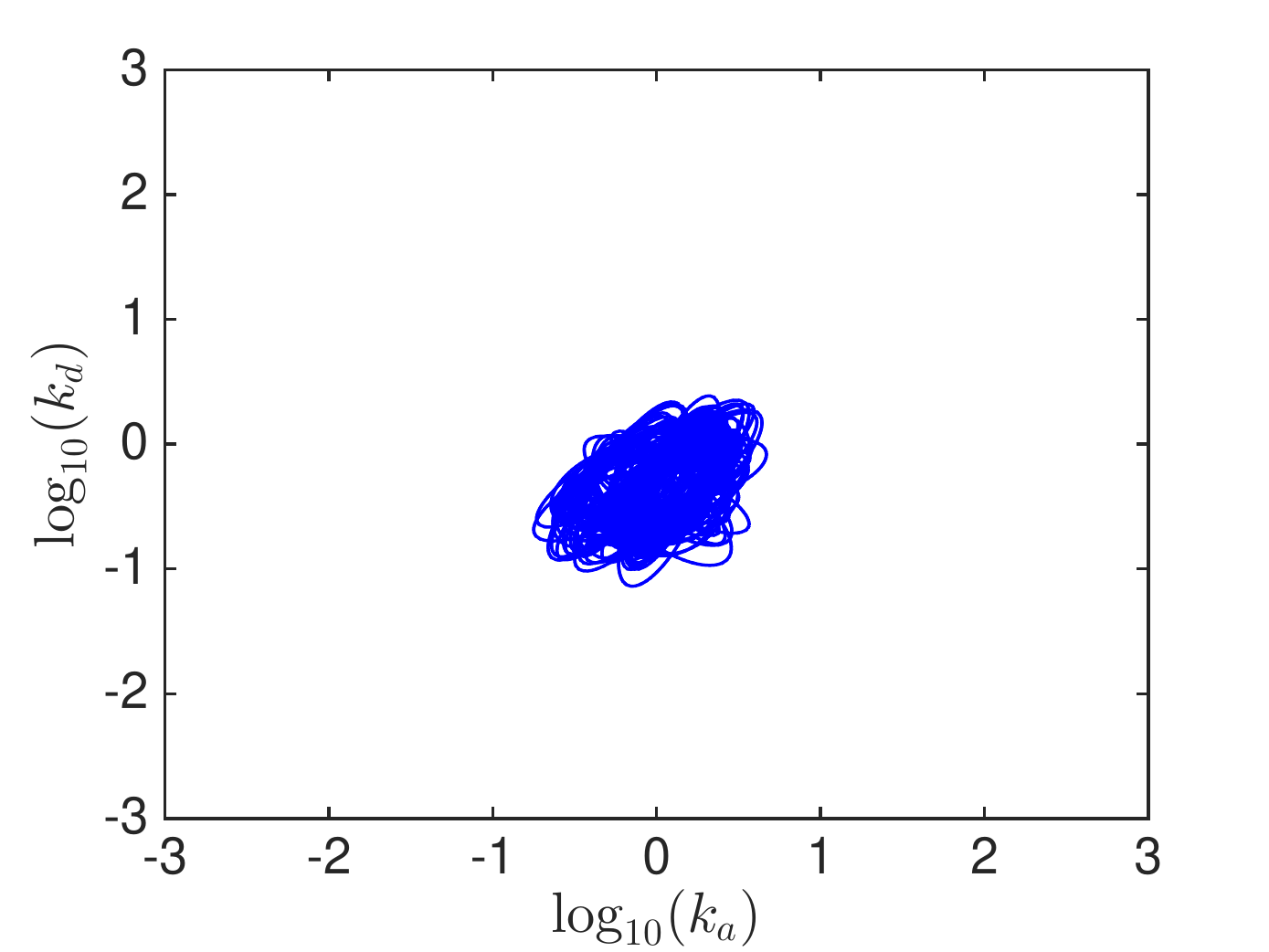}
\caption{Acceptable region}
\end{subfigure}
\caption{Average execution traces of QNSTOP (left column) based on 20 starting points and the corresponding acceptable parameter regions (right column) projected to two-dimensional domains of $k_a$ and $k_d$. (a, b) minimum distance area, (c, d) maximum log-likelihood, (e, f) approximate maximum log-likelihood. }
\label{fig:region_full_ka}
\end{figure}

\begin{figure}[!htb]
\centering
\begin{subfigure}[]{0.45\textwidth}
\centering
\includegraphics[width=1\textwidth]{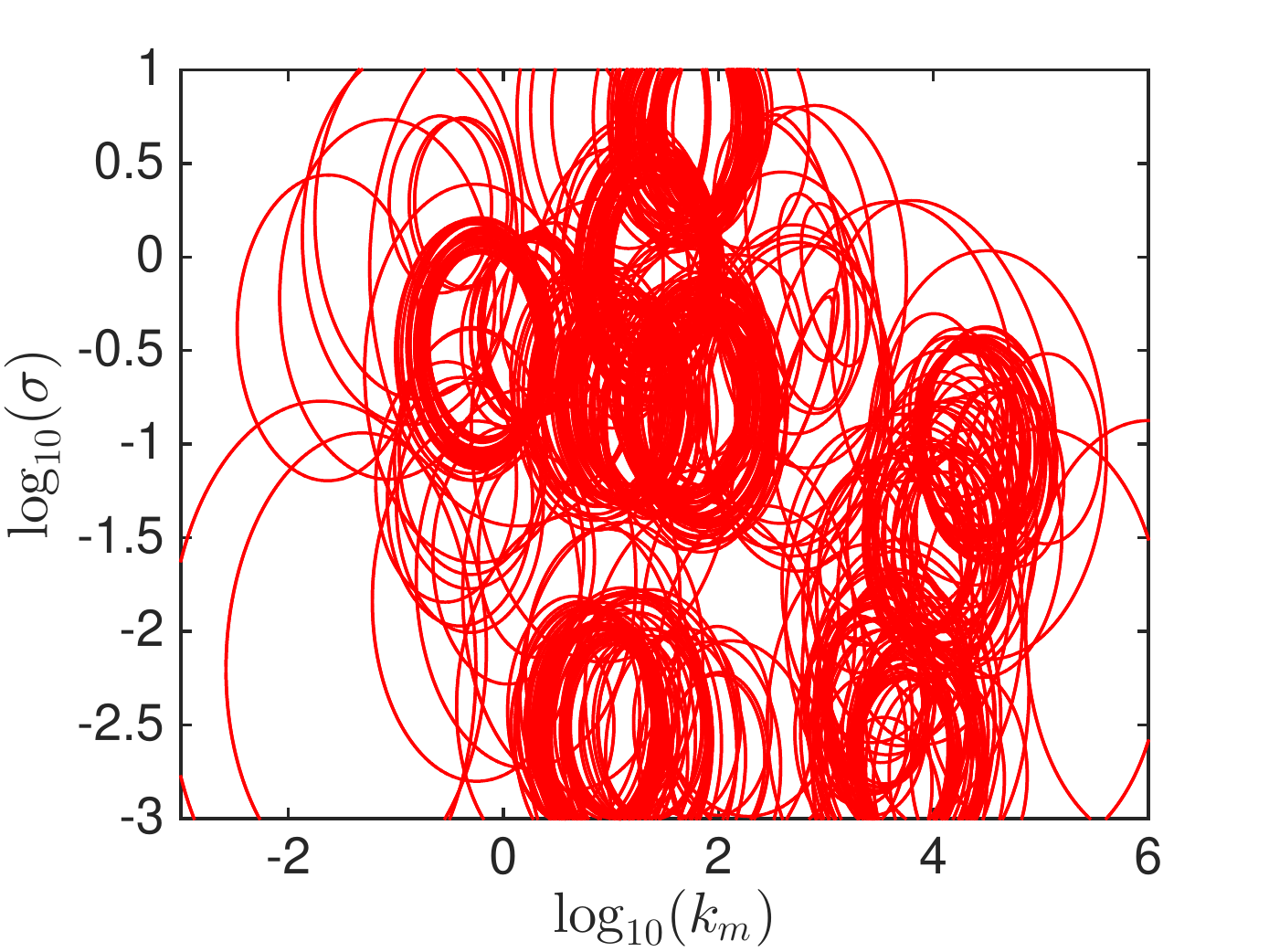}
\caption{Acceptable region for $k_m, \sigma$}
\end{subfigure}
\hspace{2em}
\begin{subfigure}[]{0.45\textwidth}
\centering
\includegraphics[width=1\textwidth]{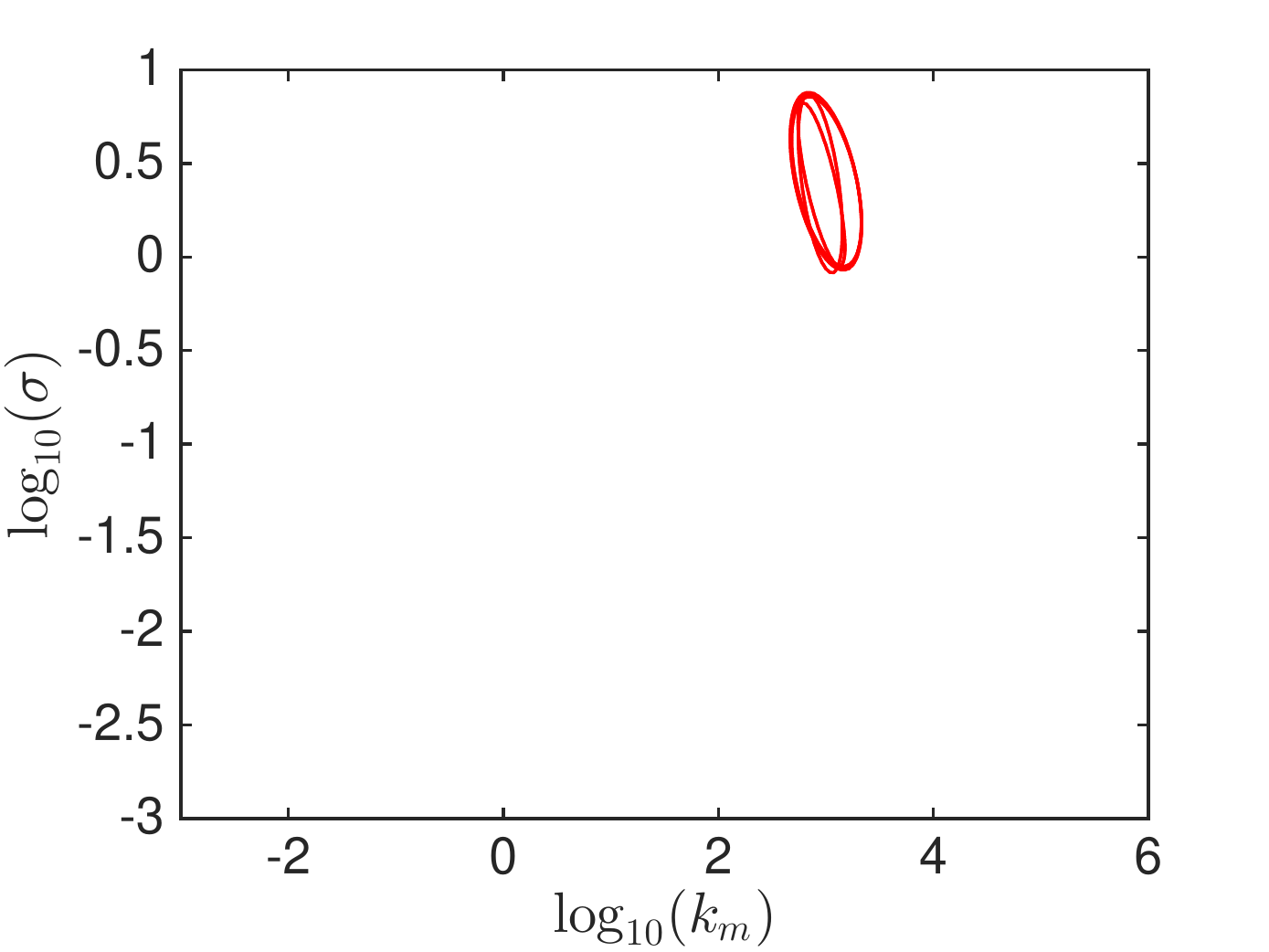}
\caption{Acceptable region for $k_m, \sigma$}
\end{subfigure}
\caption{{Acceptable parameter regions projected to two-dimensional domains
of $k_m$ and $\sigma$} from maximum log-likelihood method. (a) The population
of enzyme A is fixed at a single value. (b) 11 population levels of enzyme
A are considered in the stochastic Hill function system.}
 \label{fig:region_full_sigma}
\end{figure}

\textbf{Influence of objective functions.} Fig.~\ref{fig:region_full_ka}
presents the results of the three objective functions: minimum distance
area, maximum log-likelihood, and approximate maximum log-likelihood. For
all three methods, the acceptable regions for the ($\log_{10}(k_a)$,
$\log_{10}(k_d)$) pair are an ellipsoid shape centered at the middle of
the domain, though the size has subtle differences. While for the
($\log_{10}(k_m)$, $\log_{10}(\sigma)$) pair, the acceptable regions are
all over the domain, shown in Fig.~\ref{fig:region_full_sigma}a. This
results happens to all three methods, indicating that the system is not
sensitive to parameters $k_m$, $\sigma$. Note that here the population
level of enzyme A is fixed in this stochastic Hill function. 
The acceptable
regions of the pair ($\log_{10}(k_m)$, $\log_{10}(\sigma)$) will be
significantly different if considering multiple population levels of enzyme
A. Fig.~\ref{fig:region_full_sigma}b shows the results for an objective
function that sums over 11 population levels of enzyme A, where $[{\rm
A}]=150$, 300, 400, 520, 620, 800, 1050, 1400, 2100, 5000,
20000. This summed objective function will be explored in future work.

\begin{figure}[!htb]
\centering
\includegraphics[width=0.7\textwidth]{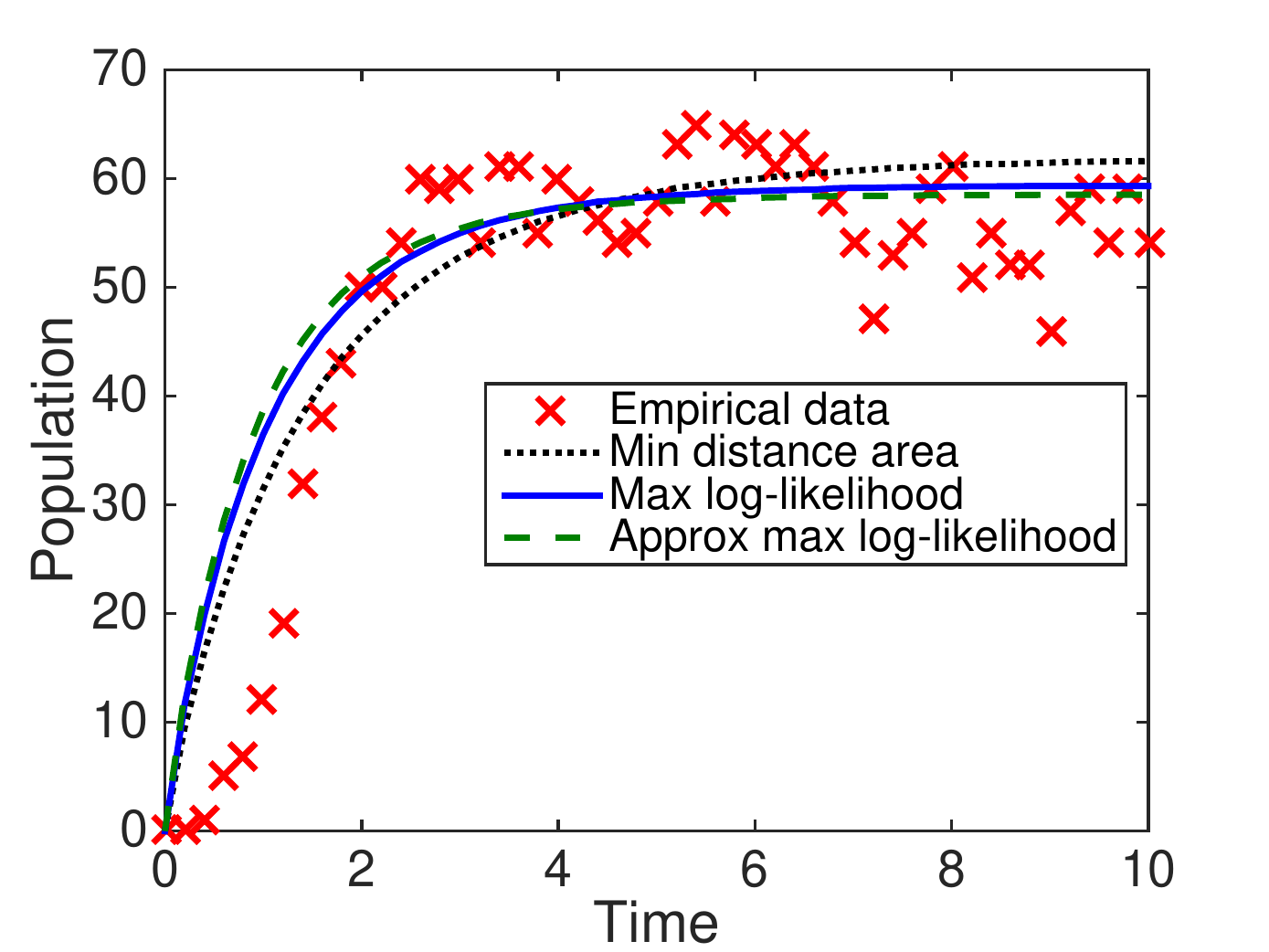}
\caption{{Average population dynamics of ${\rm B}_n$ in the stochastic Hill
function system based on 100 parameter vector values sampled from the
acceptable regions returned by minimum distance area, maximum log-likelihood,
and approximate maximum log-likelihood. The simulated empirical data (50
data points) are marked by red crosses.}}
\label{fig:avg_evol}
\end{figure}

\begin{figure}[!htb]
\centering
\begin{subfigure}[]{0.45\textwidth}
\centering
\includegraphics[width=1\textwidth]{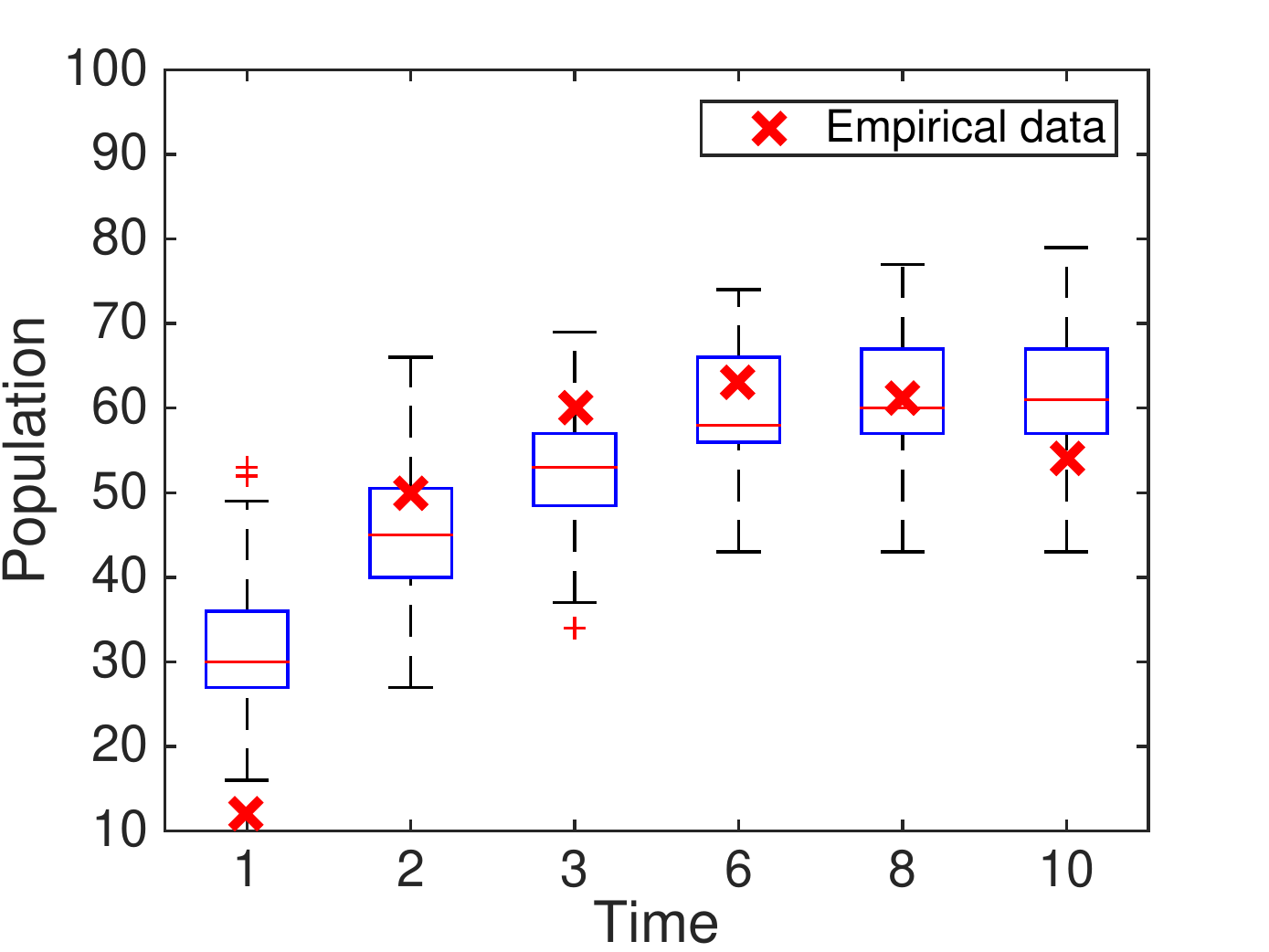}
\caption{Minimum distance area}
\end{subfigure}
\hspace{2em}
\begin{subfigure}[]{0.45\textwidth}
\centering
\includegraphics[width=1\textwidth]{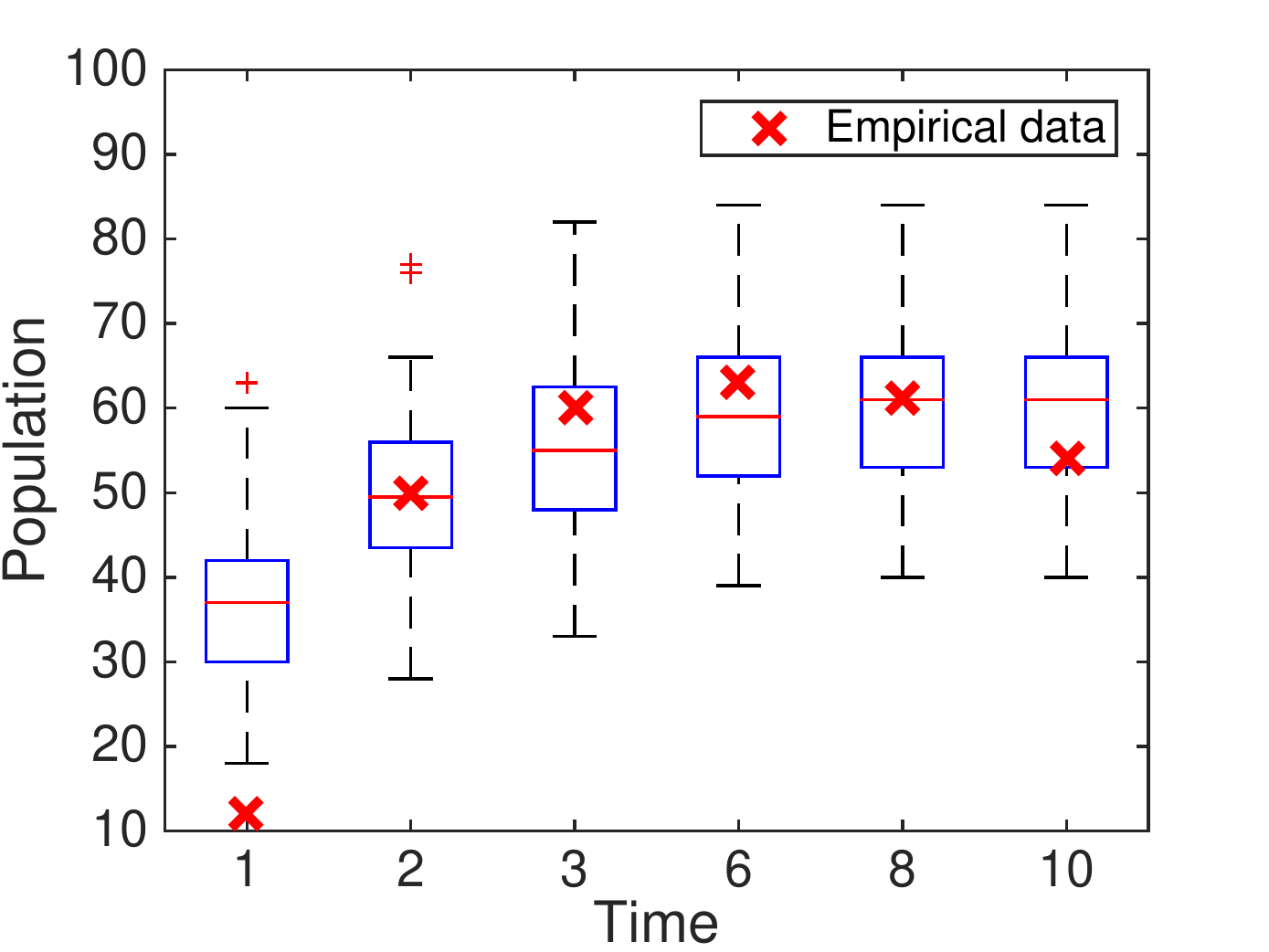}
\caption{Maximum log-likelihood}
\end{subfigure}
\begin{subfigure}[]{0.45\textwidth}
\centering
\includegraphics[width=1\textwidth]{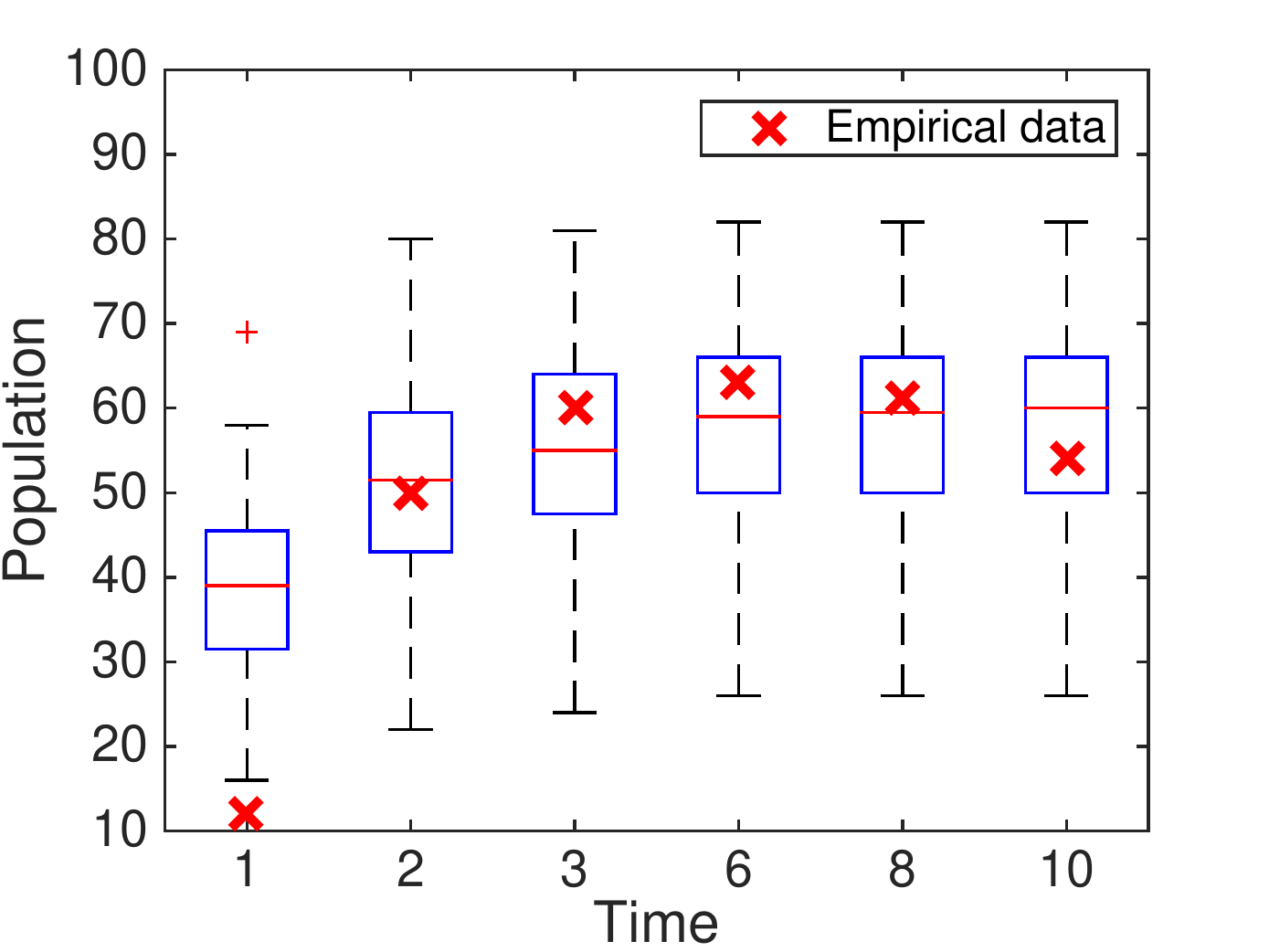}
\caption{Approximate maximum log-likelihood}
\end{subfigure}
\caption{Population distributions of ${\rm B}_n$ in the stochastic Hill
function system~\eqref{sys:hill} at time $t=1$, 2, 3, 6, 8, 10, corresponding
to 10\%, 20\%, 30\%, 60\%, 80\%, 100\% of the total simulation time, based
on 100 points sampled in each acceptable parameter region from minimum
distance area, maximum log-likelihood, and approximate maximum
log-likelihood.}
 \label{fig:evol_dist}
\end{figure}

To compare the stochastic Hill function system with the ordered distributive
enzyme-substrate phosphoryaltion process, sample 100 points from the
returned acceptable parameter region. Fig.~\ref{fig:avg_evol} shows the
average population trajectory of ${\rm B}_n$ in the stochastic Hill function
using the sampled parameter values. The average population dynamics from
all three methods match well with the empirical data.
Fig.~\ref{fig:evol_dist} further demonstrates the population distributions
of the stochastic Hill function system based on the 100 parameter vector
values sampled from the acceptable regions. Except for the significant
difference at the initial stage of the ${\rm B}_n$ transition (time $t<1$),
the empirical data falls in the 25th-75th percentile range for other stages
(time $t=2$, $t=3$, $t=6$, $t=8$, $t=10$) from maximum log-likelihood and
approximate maximum log-likelihood methods. The 25th-75th percentile range
of ${\rm B}_n$ population from minimum distance area is  narrower than that
from the other two methods.

\section{Conclusion}~\label{sec:conclusion}
This paper formulated a stochastic Hill function model for the multisite
phosphorylation mechanism, and matched SSA simulation-based empirical
population trajectory data for the ordered distributive enzyme-substrate
phosphorylation process, using three different objective functions ---
minimum distance area, maximum log-likelihood, and approximate maximum
log-likelihood. The approximate maximum log-likelihood method works well
and is applicable to large complex biochemical networks. The main
contributions are (1) an $\alpha$-$\beta$-$\gamma$ rule to find acceptable
parameter regions instead of a single best parameter vector, (2) the Hill
function system model must be matched to data with varying enzyme levels
[A], and (3) demonstrating again that QNSTOP is well-suited to stochastic
biological system model parameter estimation. Results showed that the
optimized stochastic Hill function can be used to model the switch behavior
and the steady state of the multisite phosphorylation process, while it
cannot capture the initial transition period. QNSTOP and the
$\alpha\hbox{-}\beta\hbox{-}\gamma$ rule are generally applicable to other
stochastic models. 
Meanwhile, for aforementioned first order reaction networks where 
theoretical solution or efficient modeling and simulation methods of the stochastic kinetics are available, 
one may take advantage of these theoretical solutions or reduced models and develop 
more efficient parameter estimation strategies.

\section*{Acknowledgments}
This work was partially supported by the National Science Foundation under awards CCF-1526666, MCB-1613741 and CCF-1909122.

\bibliographystyle{unsort}
\bibliography{myRef} 

\end{document}